\documentclass[a4paper]{article}

\usepackage[english]{babel}
\usepackage[utf8x]{inputenc}
\usepackage{amsmath}
\usepackage{graphicx}
\usepackage[colorinlistoftodos]{todonotes}

\usepackage[english]{babel}
\usepackage[utf8x]{inputenc}
\usepackage{amsmath}
\usepackage{amsfonts}
\usepackage{graphicx}
\usepackage{float}
\usepackage{hyperref}
\usepackage{url}
\usepackage{amsthm}

\newcommand{\dz}{\partial_z}
\newcommand{\dzbar}{\partial_{\bar z}}
\DeclareMathOperator{\tr}{Tr}
\usepackage{geometry}

\begin{document}
\title{
\begin{flushright}\ \vskip -2cm {\small {DAMTP-2014-84}}\end{flushright}
\vskip 15pt
\bf{Instanton Solutions from Abelian Sinh-Gordon and Tzitzeica Vortices}
\vskip 30pt}
\author{
Felipe Contatto$^{1,2}$, Daniele Dorigoni$^2$\\[30pt]
{\em \normalsize
$\phantom{}^1$ CAPES Foundation, Ministry of Education of Brazil, }\\[0pt] 
{\em \normalsize Bras\'ilia - DF 70040-020, Brazil. }\\[10pt] 
{\em \normalsize
$\phantom{}^2$ Department of Applied Mathematics and Theoretical Physics,}\\[0pt] 
{\em \normalsize University of Cambridge, Wilberforce Road, 
Cambridge CB3 0WA, U.K.}\\[10pt] 
{\small Email: felipe.contatto@damtp.cam.ac.uk,\ D.Dorigoni@damtp.cam.ac.uk}\\[5pt]
}
\vskip 30pt
\date{February 2015}
\maketitle
\vskip 30pt

\begin{abstract}

We study the Abelian Higgs vortex solutions to the sinh-Gordon
equation and the elliptic Tzitzeica equation. Starting from these
particular vortices, we construct solutions to the Taubes equation
with higher vortex number, on surfaces with conical singularities.

We then, analyse more general properties of vortices on such singular
surfaces and propose a method to obtain vortices on conifolds from
vortices on surfaces of revolution. We apply our method to construct
explicit vortex solutions on the Poincar\'e disk with a conical
singularity in the centre, to which we refer as the ``hyperbolic
cone".

We uplift the Abelian sinh-Gordon and Tzitzeica vortex solutions to
four dimensions and construct cylindrically symmetric, self-dual
Yang-Mills instantons on a non-self-dual (nor anti-self-dual) $4$-dimensional K\"ahler manifold with non-vanishing scalar curvature.
The instantons we construct in this way cannot be obtained via a
twistorial approach.
\end{abstract}

\newpage

\section{Introduction}

The Abelian Higgs model is a gauge field theory on a $2+1$-dimensional manifold, $\Sigma\times\mathbb R$, where $\mathbb R$ parametrizes the time and $\Sigma$ is a surface with Riemannian metric. This model describes type I and type II superconductors \cite{Abrikosov1957} depending on the value of the coupling constant, whose critical value separates both types of superconductivity. For example, the model in a non-planar geometry is physically relevant in describing thin superconductors of curved shape. In the critically coupled regime, the theory admits topological solitons called vortices, finite energy solutions to the Bogomolny equations \cite{Bogomolny76}.

The Bogomolny equations are not integrable in general, and we do not have an analytic form for the vortex profile function. Only in few lucky cases we can construct explicit solutions. In particular \cite{Witten1977,Strachan92,Manton2010}, when $\Sigma$ is a Riemann surface of constant Gauss curvature $-\frac{1}{2}$ , the Bogomolny equations reduce to the Liouville equation and analytic solutions can be found \cite{Crowdy}. Similarly, when the conformal factor on $\Sigma$ is of a very particular type, with a conical singularity at the origin, the Bogomolny equations reduce either to the sinh-Gordon or to the Tzitzeica equation \cite{MD2009}. Solutions to these equations are not known in explicit forms, but since the radial reductions of the sinh-Gordon and the Tzitzeica equations are both special cases of Painlev\'e III ODEs, we can obtain an explicit asymptotic expansion for the solutions close to the vortex centre and far away from it, without having to rely on numerical simulations \cite{Kit0,MCoy1977}.

Even if the vortex equations are non linear, it is still possible to superpose multiple vortex solutions \cite{Baptista2013}. 
This non-linear superposition rule allows us to obtain multi-vortex solutions on top of the sinh-Gordon and Tzitzeica vortices by solving an auxiliary Taubes equation on a different conically singular manifold. Singular Abelian vortex equations arise naturally as an effective tool to study vortex solutions that are invariant under the action of a symmetry group or when other type of constraints are present \cite{Baptista:2012ds}.

In \cite{Popov2009}, Popov proved that there is a one-to-one correspondence between vortices on $\Sigma$ and cylindrically symmetric instantons on $\Sigma\times S^2$. In particular, once we find an (anti-)vortex solution, we can find a solution to the (anti-)self-dual Yang-Mills, (A-)SDYM, equations on this $4$-dimensional manifold. This correspondence is a particular type of symmetry reduction of SDYM or ASDYM \cite{Masonbook,MDbook}. When the metric on $\Sigma\times S^2$ is K\"ahler and it has vanishing scalar curvature, we can use the twistor transform to construct instanton solutions. In this case, the uplifting of the vortex solutions correspond to rank-$2$ holomorphic vector bundles over the complex twistor space of $\Sigma\times S^2$.

We can apply this equivariant reduction to uplift the vortices arising from the sinh-Gordon and Tzitzeica equations and obtain instanton solutions in four dimensions. An interesting aspect of these instantons is that their backgrounds are K\"ahler manifolds of non-vanishing scalar curvature (they are not ``scalar-flat") and thus the twistor transform cannot be used.

The paper is organised as follows.
In Section \ref{secAbelianVortices} we introduce the Abelian Higgs model, the Bogomolny equations and present the vortex solutions from the sinh-Gordon and the Tzitzeica equations. Thanks to the work of Baptista \cite{Baptista2013}, in Section \ref{secSupVortices} we superpose additional vortices, in a non-linear way, on top of the sinh-Gordon and Tzitzeica solutions and construct multi-vortex solutions on particular conically singular spaces. We generalise the results of the first two sections to vortices on conifolds in Section \ref{SecCone} and, as an application of our analysis, we use the explicit analytic expression for vortices on the Poincar\'e disk to construct vortex solution on a conically singular hyperbolic space. Finally, in Section \ref{SecYM}, we briefly review the correspondence between vortices and instantons and use it to obtain cylindrically symmetric solutions of the SDYM equation by uplifting the vortices constructed in the first part.

\section{Abelian vortices} \label{secAbelianVortices}

Let us consider a Riemann surface $\Sigma$ with metric written in isothermal coordinates
$$
ds^2=\Omega(z,\bar{z}) dz d\bar z,
$$
where $z=x+iy$ is a local holomorphic coordinate. The function $\Omega:\Sigma\to \mathbb{R}^+$ is called the conformal factor of the metric.

Let $L$ be a Hermitian complex line bundle over $\Sigma$. For a certain trivialization, the Abelian $U(1)$ connection on $L$ is given by $A=A_x dx+A_y dy=A_z dz+A_{\bar{z}} d\bar z$ and its curvature is the magnetic field $B=\partial_x A_y-\partial_y A_x=2i(\dzbar A_z-\dz A_{\bar{z}})$, where we defined $\partial_z=(\partial_x-i \partial_y)/2,\partial_{\bar{z}}=(\partial_x+i \partial_y)/2$ and $A_z =(A_x-i A_y)/2,A_{\bar{z}} =(A_x+i A_y)/2$.

The potential of the critically coupled Ginzburg-Landau theory, with Bradlow parameter $\tau=1$, is
$$
V=\frac{i}{4}\int_\Sigma dz\wedge d\bar{z}\left[\Omega^{-1} B^2+\overline{D_\mu\phi} D^\mu\phi+\frac{\Omega}{4}(1-\bar\phi\phi)^2\right]\,,
$$
where the Higgs field $\phi$ is a smooth global section of $L$ and $D_\mu=\partial_\mu-iA_\mu$ is the covariant derivative. The Bogomolny equations result from completing the square in the potential $V$. They are
\begin{align}
&B=\frac{\Omega}{2}\left (1-\phi\bar\phi \right) \label{vortex1}\\
&D_{\bar{z}}\,\phi=0. \label{vortex2}
\end{align}
Vortices are defined as finite energy solutions of (\ref{vortex1})--(\ref{vortex2}). 

We set $\phi=e^{\frac{h}{2}+i\chi}$, where $h$ is a real function defined on $\Sigma$ and the phase $\chi$ is a real function defined on each open patch depending on the gauge choice. We can calculate $A_z$ and $A_{\bar z}=\overline{A_z}$ from (\ref{vortex2}) and substitute them in (\ref{vortex1}), yielding the Taubes equation
\begin{equation}\label{Taubeseq}
\Delta_0 h +\Omega\left(1-e^h\right)=0,
\end{equation}
where $\Delta_0=4\partial_z\partial_{\bar z}$ is the flat Laplacian.

The Higgs field $\phi$ vanishes at $N$ isolated points $\{Z_i\}$, the vortex locations, and precisely at these points $h$ possesses logarithmic singularities. The centres $Z_i$ are not necessarily distinct, the number of vortices, or the number of zeros of $\phi$ counted with multiplicity, is equal to the first Chern number of the bundle 
\begin{equation}
N=\frac{i}{4\pi}\int_\Sigma dz\wedge d\bar{z}\, B\,.
\end{equation}
By integrating (\ref{vortex1}) over $\Sigma$, Bradlow \cite{Bradlow1990} showed that on a surface of finite area, $N$ is bounded by
\begin{equation}\label{Bradlow}
A_{\Sigma}=\frac{i}{2}\int_{\Sigma} dz\wedge d\bar{z}\,\Omega > 4 \pi N\,,
\end{equation}
where $A_{\Sigma}$ is the area of $\Sigma$. We can interpret (\ref{Bradlow}) as saying that the effective area of a vortex is $4\pi$. 

Whenever $\vert \phi \vert =0$, $h$ has a logarithmic singularity and this implies that (\ref{Taubeseq}) is only valid away from the zeroes $\{z_i\}$ of $\vert \phi\vert$ and Taubes equation should be corrected with delta-function sources as
\begin{equation}\label{Taubeseqdelta}
\Delta_0 h+\Omega \left(1- e^h\right)=4\pi\sum^N_{i=1}\delta^2(z-z_i).
\end{equation}

For smooth and geodesically complete metrics, when all the vortices are located at the same point, i.e. $z_i=z_0$ for all $i$, $h$ can be expanded around $z = z_0$ as 
\begin{align}
h(z,\overline{z}) \sim&\notag 2N\log|z-z_0| + a(z_i,\overline{z_0}) 
+ \frac{1}{2} {\overline{b}}(z_0,\overline{z_0}) (z-z_0) + \frac{1}{2} b(z_0,\overline{z_0}) (\overline{z}-\overline{z_0}) \\
&\label{eq:hexp}
+ {\overline{c}}(z_0,\overline{z_0})(z-z_0)^2 + d(z_0,\overline{z_0})(z-z_0)(\overline{z}-\overline{z_0}) + c(z_0,\overline{z_0})(\overline{z}-\overline{z_0})^2
+ \cdots \,.
\end{align}
Apart from the leading logarithmic term, this expansion is a Taylor
series in $z-z_0$ and its conjugate. The Taubes equation (\ref{Taubeseqdelta})
requires that $d(z_0,\overline{z_0}) = - \Omega (z_0,\overline{z_0}) / 4$, but the other
coefficients shown here are not determined purely locally, but only 
from the complete $1$-vortex solution.

If we look for a circular symmetric solution of the form $h=h(r)$, with $r=\lvert z \rvert$, then (\ref{eq:hexp}) allows us to expand $h(r)$ around $r=0$ as
\begin{equation}\label{hdevelop}
h(r)\sim 2N\ln r+a-\frac{\Omega(0)}{4} r^2+O(r^4).
\end{equation}

We underline that equation (\ref{hdevelop}) is true only when the metric is smooth and geodesically complete, in particular we require $\Omega(0)$ to be well-defined. In what follows we will give examples of metrics with conical singularities at the origin $r=0$; the vortex solutions to Taubes equation with these singular conformal factors will have asymptotic series in the origin in fractional powers of $r$, also known as Puiseux series.

We will be working mainly with surfaces of revolution, which admit $z\mapsto ze^{i \varphi}$ as a one parameter group of isometries. We point out that since there is a unique solution to (\ref{Taubeseqdelta}) once we fix all the vortex positions $z_i$ and since (\ref{Taubeseqdelta}) is invariant under isometries of the manifold, vortices at the origin of a surface of revolution are necessarily rotationally invariant. This translates the intuitive fact that, in a surface of revolution, there is no preferred radial direction.

Noticing that $h$ vanishes at $r\to\infty$, equation (\ref{Taubeseqdelta}) reduces to a Bessel equation whose solution has the asymptotic behavior
\begin{equation}\label{hdevelopinfty}
h(r)\sim\frac{\Lambda}{\sqrt r}e^{-\sqrt{\Omega_{as}}r}, 
\end{equation}
where $\Omega_{as}=\lim_{r\to\infty}\Omega$ and $\Lambda$ is a constant, called the vortex strength.

Given the two asymptotic forms (\ref{hdevelop}) --(\ref{hdevelopinfty}), Taubes equation uniquely determines the constants $a$ and $\Lambda$ but, since generically an explicit solution is not known, they have to be computed numerically in most situations.
 In the flat case, $\Omega=1$, it is possible \cite{VegaS1976} to relate the two asymptotic expansions and effectively reduce the problem of solving Taubes equation to a system of transcendental algebraic equations relating $a$ and $\Lambda$.
A particularly special case is when $\Sigma$ is a hyperbolic surface of constant Gauss curvature $-\frac{1}{2}$, for which we can obtain an exact solution to the Taubes equation by reducing the problem to a Liouville equation \cite{Witten1977,Strachan92,Manton2010}. 
Other two integrable cases, focus of the present work, are when the Taubes equation reduces to the sinh-Gordon or to the Tzitzeica equations \cite{MD2009}. When the conformal factor is chosen in a very peculiar way, we can reduce the radially symmetric Taubes equation to particular cases of Painlev\'e III ODEs, and even if an explicit solution is not known for all $r$, we can recover analytically the two asymptotic forms (\ref{hdevelop})--(\ref{hdevelopinfty}) for $h$, and the connection formulas for $a$ and $\Lambda$. 

\subsection{The Sinh-Gordon Vortex}\label{SecSGVortex}

In this Section, following \cite{MD2009}, we describe how to reduce Taubes equation to the sinh-Gordon equation and present the solution in the two asymptotic regimes (\ref{hdevelop}--\ref{hdevelopinfty}). 

Let us consider the vortex equations on the surface $\Sigma=\mathbb C$ whose metric, in isothermal coordinates, has the conformal factor $\Omega=e^{-h/2}$ and reads
$$
g_{\Sigma}=e^{-h(z,\bar{z})/2}dzd\bar z.
$$
Note that the vortex field is included in the Riemannian data of the background.

In this case, (\ref{Taubeseq}) becomes the elliptic sinh-Gordon equation 
\begin{equation}\label{Sinh}
\Delta_0 (h/2)=\sinh(h/2).
\end{equation}

Looking for a solution with rotational symmetry, we assume that $h$ depends only on the radial component $r=\lvert z \rvert$. This implies that the vortex position, the point where the Higgs field $\phi$ vanishes, must be the origin.

An $N$ sinh-Gordon vortex would be a solution to (\ref{Sinh}) with a logarithmic singularity, close to $r\sim0$, of the form $h\sim 2 N \log r$, with $N>0$ integer, and such that $h\to 0$ for $r \to \infty$.
As in (\ref{Taubeseqdelta}) the logarithmic singularity in $h$ corresponds to the zero of $\vert \phi \vert$ and it adds a delta function singularity on the right hand side of (\ref{Sinh}).

It is possible \cite{MD2009} to map equation (\ref{Sinh}) to a particular type of Painlev\'e III ODE, with parameters $(0,0,1,-1)$, the requirements that the vortex solution has a logarithmic singularity at the origin and that $h$ has to vanish for $r\to \infty$, together with the Painlev\'e property  \cite{MCoy1977}, fix uniquely the asymptotic forms of the solution.
There is a unique solution to (\ref{Sinh}) yielding a vortex solution \cite{MD2009}. This sinh-Gordon vortex has $N=1$ and for $r\sim 0$ has the asymptotic form
\begin{equation}\label{hsmallr}
h_{sG}(r)\sim2\ln(r)+4\ln\beta_{sG}-\frac{r}{\beta_{sG}^2}+O(r^2),
\end{equation}
where $\beta_{sG}=2^{-3/2}\frac{\Gamma(1/4)}{\Gamma(3/4)}\approx 1.046$, and all higher orders are fixed in terms of $\beta_{sG}$.
The asymptotic solution for $r \to \infty$ is also uniquely determined
\begin{equation}\label{hlarger}
h_{sG}(r)\sim -\Lambda_{sG} K_0(r),
\end{equation}
where $K_n$ denotes the modified Bessel functions of the second kind, which decay exponentially with $r$ precisely as (\ref{hdevelopinfty}), and the sinh-Gordon vortex strength is denoted by $\Lambda_{sG}=8\lambda\sim 1.80$, where $\lambda=\frac{\sqrt 2}{2\pi}$.

We notice in the expansion of $h_{sG}$ close to the origin, a linear term in $r$, or equivalently $\lvert z \rvert$, which should not be present according the expansion (\ref{hdevelop}). The reason for this is the presence of a pole in the conformal factor of the metric (\ref{metricrsmall1}) at $r\sim0$, the metric is not geodesically complete at the origin and the expansion (\ref{hdevelop}) does not apply.
The metric close to the origin takes the form
\begin{equation}\label{metricrsmall1}
g_{\Sigma}\sim \,\frac{1}{r\beta_{sG}^2}(dr^2+r^2d\theta^2)\,.
\end{equation}
The change $\rho=\sqrt{r}$ of the radial coordinate shows that, close to the origin, $\Sigma$ possesses a flat metric 
$$
g_{\Sigma}\sim \frac{4}{\beta_{sG}}(d\rho^2+\frac{1}{4}\rho^2d\theta^2)\,,
$$
which presents a conical singularity with deficit angle $\pi$. 

The cone is graphically visible by performing an isometric immersion of the surface $\Sigma$ into $\mathbb R^3$ 
$$
re^{i\theta}\in\Sigma \mapsto (X(r,\theta),Y(r,\theta),Z(r,\theta))=\left(\sqrt{r}\cos\theta,\sqrt{r}\sin\theta,\sqrt{3r}\right)\in\mathbb R^3
$$
whose image satisfies the equation of a cone with aperture $\pi/3$
\begin{equation}\label{Cone}
Z=\sqrt{3(X^2+Y^2)}.
\end{equation}

Note that the linear term $-r/\beta_{sG}^2$ in (\ref{hsmallr}), when expressed using the coordinate $\rho=\sqrt{r}$, takes precisely the form $-\rho^2\Omega(0)/4 $ dictated by our initial expansion (\ref{hdevelop}). It is only by using the coordinate $\rho$, for which the metric is flat and $\Omega(0)$ well defined and non-vanishing, that we can recover (\ref{hdevelop}) from (\ref{hsmallr}).
This situation will arise whenever our background metric has a conical singularity and we insist in inserting a vortex exactly at the tip of the cone: the right coordinates to use are the ones for which the metric is flat, despite the angular variable not having periodicity $2\pi$, in this way we will recover precisely the expansion (\ref{hdevelop}), see Section \ref{SecCone} for more details.

Far from the origin, the metric is perfectly smooth and takes the form
$$
g_{\Sigma}\sim e^{4\lambda K_0(\lvert z\rvert)}dz d\bar z \,,
$$
which means that for large $r$ the metric is actually flat since for $r\to\infty,\, K_0(r)\sim\sqrt{\frac{\pi}{2r}}e^{-r}$.

\subsection{The Tzitzeica Vortex}\label{SecTTVortex}

A similar analysis of the Taubes equation can be carried along for vortices coming from the Tzitzeica equation and analytic asymptotic solutions can be obtained in a similar fashion \cite{MD2009}. 

Let us consider the vortex equations on the surface $\Sigma=\mathbb C$ whose metric has the conformal factor $\Omega=e^{-2h/3}$. The metric in isothermal coordinates reads
$$
g_{\Sigma}=e^{-2h(z,\bar{z})/3}dzd\bar z.
$$
Once again the Riemannian background metric is fixed by the vortex profile function on this particular metric.

In this case, (\ref{Taubeseq}) becomes the elliptic Tzitzeica equation 
\begin{equation}\label{TT}
\Delta_0 u+\frac{1}{3} \left( e^{-2 u} - e^{u}\right)=0\,,
\end{equation}
where $ h = 3 u$.

A story similar to the one presented in the previous Section can be repeated for the Tzitzeica vortex \cite{MD2009}.
We can map (\ref{TT}) to a Painlev\'e III ODE, this time with parameters $(1,0,0,-1)$. The requirements that close to the origin, $h$ has a $2 N \log r$ singularity, with $N$ integer, and then vanishes asymptotically for $r\to \infty$, together with the Painlev\'e property \cite{Kit0}, fix uniquely the solution.
As in the sinh-Gordon case, there is a unique Tizteica vortex, which also has vortex number $N=1$, and, for $r\sim 0$, takes the asymptotic form
\begin{equation}\label{hsmallrTT}
h_{TT}(r)=3 u (r)\sim2\ln(r)+\beta_{TT}-\frac{9\,e^{-2 \beta_{TT}/3} }{4} r^{2/3}+O(r^{4/3}),
\end{equation}
where all the higher order terms are fixed in terms of $\beta_{TT}$, which can be read off from equation (19) of \cite{Kit0}:
\begin{equation*}
\beta_{TT}=3 \log\left[-\frac{3^{\nu+1}}{12^{1/3}}\frac{\Gamma(\frac{1}{2}+\frac{1}{6}\nu)\Gamma(\frac{\nu}{3})}{\Gamma(\frac{1}{2}-\frac{1}{6}\nu)\Gamma(-\frac{\nu}{3})} \right],
\end{equation*}
where $\nu=3\left( 1-\frac{p}{\pi}\right)$ and $p$ has to be set to $8\pi/9$, so $\beta_{TT}\approx 0.864$.

Similarly, for $r\gg 1$
\begin{equation}\label{hlargerTT}
h_{TT}(r)\sim \frac{6 \sqrt{3}}{\pi} \left(\cos p +\frac{1}{2}\right)K_0( r )= -\Lambda_{TT} \,K_0(r)\,,
\end{equation}
by substituting $p=8\pi/9$ we obtain that the Tzitzeica vortex strength is $\Lambda_{TT}\approx 1.45 $.

As noticed before for the sinh-Gordon vortex, also in this case the expansion for $h_{TT}$ close to the origin is not of the form (\ref{hdevelop}). Instead, it has a power series in $r^{2/3}$. The reason is once again a conical singularity for the metric at $r\sim0$.
The metric close to the origin takes the form
\begin{equation}\label{metricrsmall}
g_{\Sigma_0}\sim \frac{e^{-2\beta_{TT}/3}}{r^{4/3}}(dr^2+r^2d\theta^2)\,.
\end{equation}
With the change of variables $\rho=r^{1/3}$, we see that $\Sigma$ possesses a flat metric close to the origin
$$
g_{\Sigma}\sim 9 \,e^{-2\beta_{TT}/3}(d\rho^2+\frac{1}{9}\rho^2d\theta^2)\,,
$$
with a conical singularity with deficit angle $4 \pi/3$. The cone is embeddable into $\mathbb R^3$ as
$$
re^{i\theta}\in\Sigma \mapsto (X(r,\theta),Y(r,\theta),Z(r,\theta))=\left(r^{1/3}\cos\theta,r^{1/3}\sin\theta,\sqrt{8}\,r^{1/3}\right)\in\mathbb R^3
$$
whose image satisfies the equation of a cone with aperture $2 \cot^{-1}\sqrt{8}\,$,
\begin{equation}\label{Cone}
Z=\sqrt{8(X^2+Y^2)}.
\end{equation}

Also in the Tzitzeica case, if we use the right coordinate $\rho = r^{1/3}$, for which the metric is flat and $\Omega(0)$ well defined, we can rewrite the term $-\frac{9\,e^{-2 \beta_{TT}/3} }{4} r^{2/3}$ in (\ref{hsmallrTT}) as $-\rho^2 \Omega(0)/4$ and recover the expansion (\ref{hdevelop}).

As for the sinh-Gordon vortex, in the Tzitzeica case as well, we have a smooth and flat metric far from the origin
$$
g_{\Sigma}\sim e^{-\Lambda_{TT} \,K_0(r)}(dr^2+r^2d\theta^2)\,.
$$

\section{Superposition of vortices}\label{secSupVortices}

In this Section we briefly review a non-linear rule \cite{Baptista2013} for superposing vortices in order to create higher vortex number solutions on $(\Sigma,g)$ by solving instead a lower vortex number Taubes equation on a modified background $(\tilde\Sigma,\tilde g)$.

Let us suppose that $h$ satisfies the Taubes equation (\ref{Taubeseqdelta}) on $(\Sigma,g)$, with vortex number $N$ and vortex centres $\{Z_i\}$, and that $\tilde h$ satisfies a second Taubes equation
\begin{equation*}
\Delta_0 \tilde h +\tilde\Omega\left( 1- e^{\tilde h} \right)=4\pi\sum_{j=1}^M \delta^2\left(z-\tilde Z_j\right) ,
\end{equation*}
where $\tilde\Omega(z,\bar{z}) =e^{h(z,\bar{z})}\Omega(z,\bar{z})$ is the conformal factor of a degenerate metric, vanishing at $\{Z_i\}$. We call $\tilde\Sigma$ the surface with metric $\tilde{g}=\tilde\Omega\, dzd\bar z$.

Now, it is straightforward to verify the identity
\begin{equation*}
\Delta_0 \left(\tilde h+h\right)+\Omega\left(1- e^{\tilde h+h}\right)=4\pi\sum_{i=1}^N \delta^2\left(z-Z_i\right)+4\pi\sum_{j=1}^M \delta^2\left(z-\tilde Z_j\right),
\end{equation*}
which shows that $h+\tilde h$ satisfies the Taubes equation on $(\Sigma,g)$, with vortex number $N+M$ and vortex locations at $\{Z_i\}\cup\{\tilde Z_j\}$.

So, as a non-linear rule for superposing vortices, instead of looking for a $N+M$ vortex on $(\Sigma,g)$ we can look for a $M$ vortex solution $\tilde h$ on $(\tilde\Sigma,\tilde g)$, with conformal factor $\tilde \Omega= e^h \,\Omega$. The combination $h+\tilde h$ is now the vortex solution on $(\Sigma,g)$ we were looking for. Note that generically, even if $(\Sigma,g)$ is smooth and geodesically complete, $(\tilde\Sigma,\tilde g)$ will not be so: the logarithmic singularities of $h$ will induce conical singularities in $\tilde g$.

\subsection{Multi-vortices from the sinh-Gordon and Tzitzeica vortices}
\label{sec:mvSGTT}

Let us apply now the superposition rule to obtain multi-vortex solutions on top of the sinh-Gordon vortex $h_{sG}$, which is defined on the surface $\Sigma=\mathbb C$ with metric $g=e^{-h_{sG}(z,\bar{z})/2}dzd\bar z$.

Firstly, we Weyl rescale the metric $g$ by $\lvert \phi\rvert^2=e^{h_{sG}}$ to find the metric $\tilde g=e^{h_{sG}} g$ on the surface $\tilde\Sigma$. In the limits $r\to 0$ and $r\to\infty$, respectively, it is given by
\begin{align}
\tilde g&\label{eq:mvSGmetric}\sim  \beta^2_{sG} r\left(dr^2+r^2d\theta^2\right)=\frac{4 \beta^2_{sG}}{9}\left(d\tilde\rho^2+\frac{9}{4}\tilde\rho^2d\theta^2\right)  &(r\to 0)\\
\tilde g&\sim e^{-4\lambda K_0(r)}\left(dr^2+r^2d\theta^2\right) &(r\to \infty)
\end{align}
where $\tilde\rho=r^{3/2}$.

\begin{center}
\begin{figure}[t]\centering{
\includegraphics[scale=0.25]{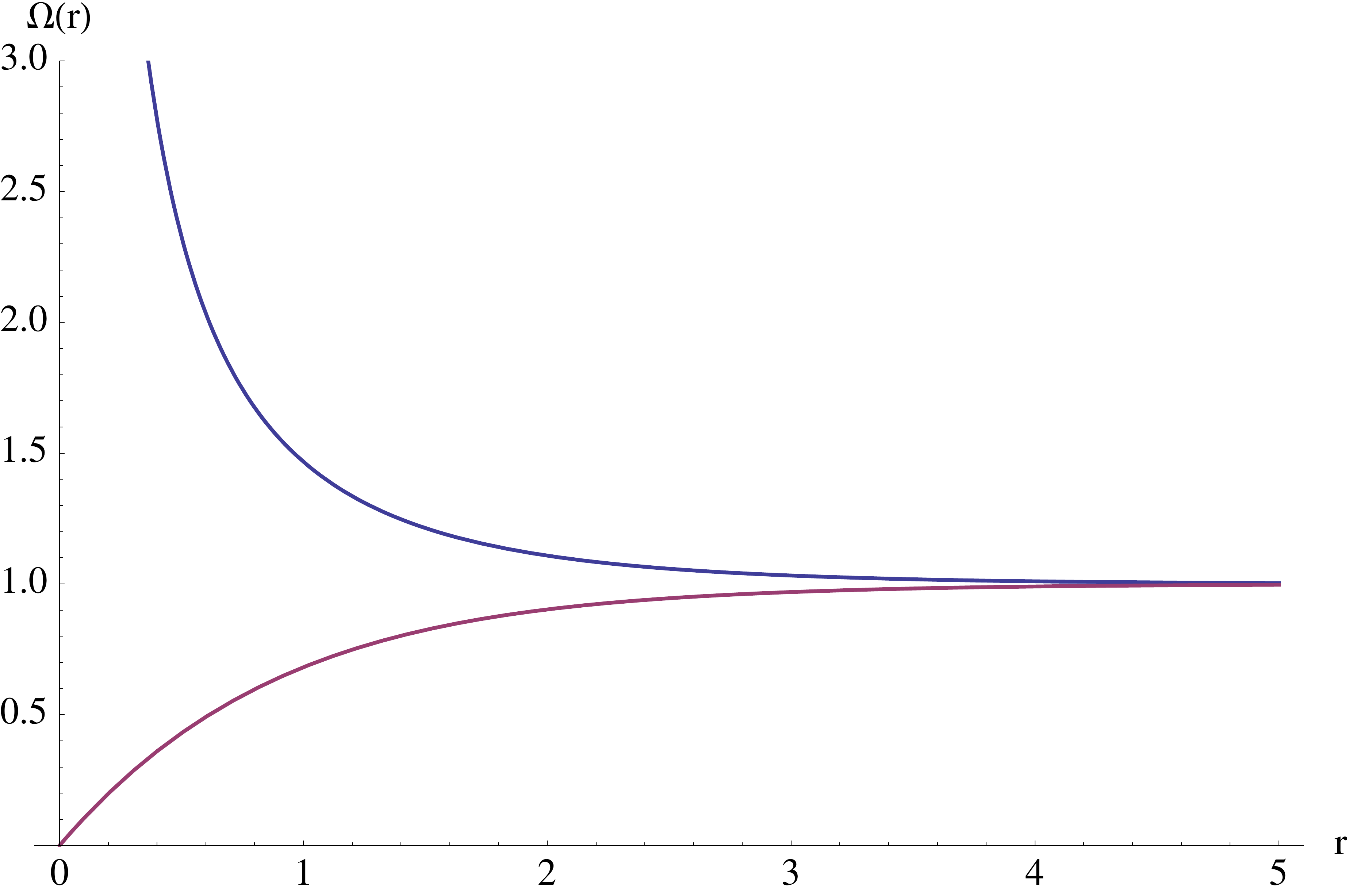}
\caption{The conformal factor $\Omega=e^{-h_{sG}/2}$ for the sinh-Gordon vortex, upper plot, and the rescaled one $\tilde \Omega = e^{h_{sG}} \,\Omega=e^{h_{sG}/2}$, lower plot.}
\label{Figure:metricSG}}
\end{figure}
\end{center}
The two conformal factors, $\Omega$ and $\tilde \Omega$, are shown in Figure \ref{Figure:metricSG}. Note that $\Omega$ diverges in the origin as $r^{-1}$ while $\tilde \Omega$ goes to zero as $r$, both factors tend to $1$ for large $r$ .
In a neighbourhood of the origin, $\tilde\Sigma$ looks like a cone in Minkowskian $\mathbb R^{2+1}$, as explicitly shown by the isometry
$$
re^{i\theta}\in\tilde\Sigma \mapsto (\tilde X(r,\theta),\tilde Y(r,\theta),\tilde Z(r,\theta))=\left(r^{3/2}\cos\theta, r^{3/2}\sin\theta,\frac{\sqrt{5}}{3} r^{3/2}\right)\in\mathbb R^{2+1}.
$$
It is indeed an isometry as $d\tilde X^2+d\tilde Y^2-d\tilde Z^2= r\left(dr^2+r^2d\theta^2\right)$.

In order to find a multi-vortex solutions with $N+1$ vortices located at the origin of $\Sigma$, we could try to solve the Taubes equation (\ref{Taubeseq}) with conformal factor $\Omega=e^{-h_{sG}/2}$, with $h$ given by (\ref{hsmallr}) and (\ref{hlarger}), however, as we have seen in Section \ref{SecSGVortex}, $\Omega$ is actually diverging for $r\to 0$.  To by-pass the complications of an ill-defined conformal factor we can use the superposition rule just explained. Instead of looking for an $N+1$ vortex solution on $(\Sigma,g)$ we study an $N$ vortex problem on $\tilde\Sigma$ whose metric has the conformal factor $\tilde\Omega=e^{h_{sG}}\Omega=e^{h_{sG}/2}$ which is well-defined at the origin, $\tilde\Omega(0)=0$. 
The problem of finding $N+1$ vortices at the origin in $(\Sigma,g)$ reduces to
\begin{align}
&\Delta_0 h_{sG}= 2\sinh\left( \frac{h_{sG}}{2}\right) +4\pi\delta^{2}(z)\, , \label{system1}\\
&\Delta_0 \tilde h +e^{h_{sG}/2}\left(1 -e^{\tilde h}\right)=4 \pi N \delta^{2}(z)\,, \label{system2}
\end{align}
where $h_{sG}$ satisfies (\ref{hsmallr}) and (\ref{hlarger}), while $\tilde h$ has the asymptotic expansions 
\begin{align}
\tilde h(r)&\label{eq:mvSG}\sim2N\ln r+\tilde a-\frac{\beta_{sG}^2}{9} r^3 +O(r^4) &(r\to 0)\\
\tilde h(r)&\sim \tilde\Lambda\, K_0(r) &(r\to \infty)\,,
\end{align}
where all the higher orders are uniquely determined in terms of $\tilde a$ (or equivalently $\tilde \Lambda$) and $\beta_{sG}$.
Note that both $\tilde a$ and $\tilde \Lambda$ are constants in $r$ but depend actually on the vortex number $N$.
Through a numerical analysis of the system (\ref{system1}--\ref{system2}), see the Appendix for more details, we obtained these constants  $\tilde a$ and $\tilde \Lambda$ for various vortex numbers $N$. Note, by the way, that since the area of $\Sigma$ is infinite, the vortex number $N$ is not limited by the Bradlow inequality (\ref{Bradlow}).
For the $N=1$ case, our numerical analysis gave $\tilde a=-1.43 ,\, \tilde \Lambda =-4.69 $.

Once again the expansion (\ref{eq:mvSG}) is not of the form (\ref{hdevelop}), because the metric, due to the conical singularity at the origin, is not geodesically complete, even if this time the conformal factor is not diverging.
From (\ref{eq:mvSGmetric}), we discover that if we use the coordinate $\tilde\rho=r^{3/2}$, the metric $\tilde{g}$ becomes flat and with a non-vanishing conformal factor $\tilde \Omega (0) =4 \beta^2_{sG}/9$. The term $-r^3 \beta_{sG}^2/9 $ in (\ref{eq:mvSG}) is precisely $-\tilde\rho^2\tilde\Omega(0)/4$ as expected from (\ref{hdevelop}).

The story can be repeated verbatim for the Tzitzeica vortex $h_{TT}$, defined on the surface $\Sigma=\mathbb C$ with metric $g=e^{-2h_{TT}(z,\bar{z})/3}dzd\bar z$.
Rescaling the metric $g$ by $\lvert \phi\rvert^2=e^{h_{TT}}$ we find the new conformal factor $\tilde \Omega =e^{h_{TT}} \Omega=e^{h_{TT}/3}$ of a new surface $\tilde\Sigma$. In the limits $r\to 0$ and $r\to\infty$, respectively, this metric is given by
\begin{align}
\tilde g&\label{eq:mvTTmetric}\sim e^{\beta_{TT}/3} r^{2/3} \left(dr^2+r^2d\theta^2\right)=\frac{9 e^{\beta_{TT}/3}}{16}\left(d\tilde\rho^2+\frac{16}{9}\tilde\rho^2d\theta^2\right)  &(r\to 0)\\
\tilde g&\sim e^{ -\Lambda_{TT}\,K_0(r)}\left(dr^2+r^2d\theta^2\right) &(r\to \infty)
\end{align}
where $\tilde\rho=r^{4/3}$.

\begin{center}
\begin{figure}[t]
\centering{
\includegraphics[scale=0.25]{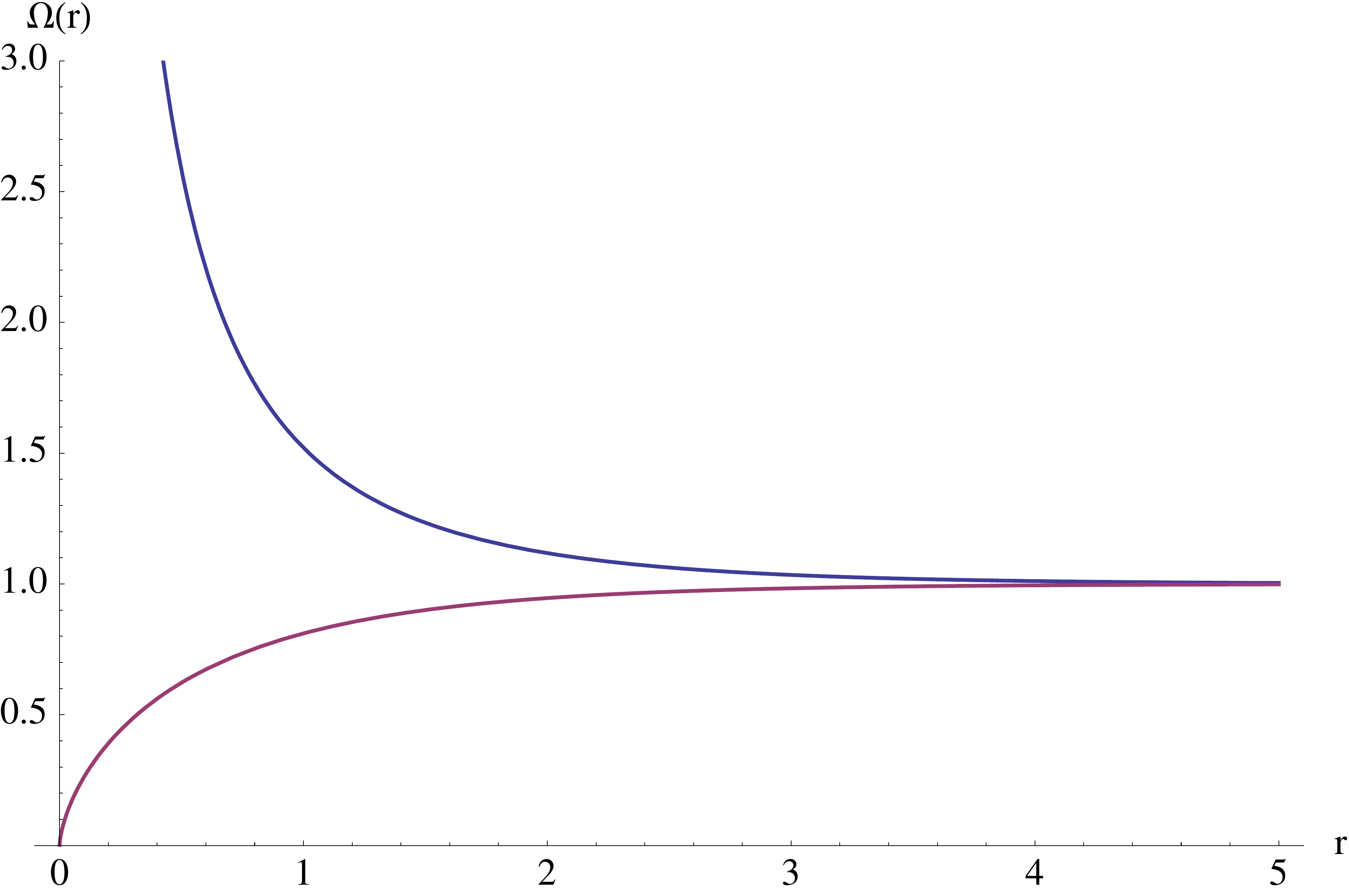}
\caption{The conformal factor $\Omega=e^{-2 h_{TT}/3}$ for the Tzitzeica vortex, upper plot, and the rescaled one $\tilde \Omega = e^{h_{TT}} \,\Omega=e^{h_{TT}/3}$ , lower plot.}
\label{Figure:metricTT}}
\end{figure}
\end{center}
The two conformal factors, $\Omega$ and $\tilde \Omega$, are shown in Figure \ref{Figure:metricTT}, note that $\Omega$ diverges in the origin as $r^{-4/3}$ while $\tilde \Omega$ goes to zero as $r^{2/3}$, both factors tend to $1$ for large $r$ .
In a neighbourhood of the origin, $\tilde\Sigma$ looks like a cone in Minkowskian $\mathbb R^{2+1}$, as explicitly shown by the isometry
$$
re^{i\theta}\in\tilde\Sigma \mapsto (\tilde X(r,\theta),\tilde Y(r,\theta),\tilde Z(r,\theta))=\left(r^{4/3}\cos\theta, r^{4/3}\sin\theta,\frac{\sqrt{7}}{4} r^{4/3}\right)\in\mathbb R^{2+1}.
$$
It is an isometry since $d\tilde X^2+d\tilde Y^2-d\tilde Z^2= r^{2/3}\left(dr^2+r^2d\theta^2\right)$.

As we did before, instead of studying the problem of finding $N+1$ vortices, located at the origin in $(\Sigma,g)$, we first solve for the Tzitzeica vortex and then we look for an $N$ vortex solution on $(\tilde\Sigma,\tilde g)$:
\begin{align}
&\Delta_0 h_{TT} +e^{ - 2 h_{TT} /3 } \left(1-e^{h_{TT}}\right)=4 \pi \delta^{2}(z)  , \label{system1TT}\\
&\Delta_0 \tilde h +e^{h_{TT}/3}\left( 1 - e^{\tilde h}\right) =4 \pi N \delta^{2}(z) \,,\label{system2TT}
\end{align}
where $h_{TT}$ satisfies (\ref{hsmallrTT}) and (\ref{hlargerTT}), while $\tilde h$ has the asymptotic expansions 
\begin{align}
\tilde h(r)&\label{eq:mvTT}\sim2N\ln r+\tilde a-\frac{9 \,e^{\beta_{TT}/3}}{64} r^{8/3} +O(r^{10/3}) &(r\to 0)\\
\tilde h(r)&\sim  \tilde\Lambda \,K_0(r)&(r\to \infty)\,,
\end{align}
where, once again, all the higher orders are uniquely determined in terms of $\tilde a$ (or equivalently $\tilde \Lambda$) and $\beta_{TT}$.
For the $N=1$ case, our numerical analysis gave $\tilde a=-1.28 ,\, \tilde \Lambda = -4.18$.

Even in this case, if we take (\ref{eq:mvTT}) and rewrite it using the flattening coordinate $\tilde\rho=r^{4/3}$, for which the metric (\ref{eq:mvTTmetric}) has a non-vanishing conformal factor in the origin $\tilde \Omega(0)=9 e^{\beta_{TT}/3}/16$, we recover precisely the term $-\tilde \rho^2\Omega(0)/4$ as expected from the expansion (\ref{hdevelop}).

\section{Vortices on conically singular spaces}\label{SecCone}

In this Section we derive the asymptotic expansion for the vortex profile function $h$, close to the vortex centre, when the background surface $(\Sigma,g)$ is a cone.
For simplicity we will assume that the metric $g$ takes the form
\begin{equation}
g= r^{2\alpha} \left(dr^2+r^2d\theta^2 \right) = 
\vert z \vert ^{2\alpha} dz\,d\bar{z}\,,\label{eq:metricConer}
\end{equation}
with $1+\alpha>0$ and for all $r\in\mathbb{R}^+$, not just close to $r\sim 0 $ as before.
With the change of variables $\rho = r^{1+\alpha}$ the metric becomes flat
\begin{equation}
g = \frac{1}{(1+\alpha)^2} \left( d\rho^2 + \rho^2 (1+\alpha)^2 d\theta^2\right)= dZ\, d\bar{Z}\,,\label{eq:metricCone}
\end{equation}
where $Z= z^{1+\alpha}$, but the new angle variable $\Theta = (1+\alpha)\, \theta$ is now periodic with period $2\pi \,(1+\alpha)$, denoting precisely a conical singularity with deficit angle $2\pi\alpha$, embeddable as a cone in $\mathbb{R}^3$ for $-1< \alpha<0$ or a cone in $\mathbb{R}^{2+1}$ for $\alpha>0$.
For example in the sinh-Gordon case (\ref{metricrsmall1}) $\alpha=-1/2$ while in the multi sinh-Gordon case (\ref{eq:mvSGmetric}) $\alpha = 1/2$, similarly for the Tzitzeica vortex (\ref{metricrsmall}) $\alpha=-2/3$ and in the multi Tzitzeica vortex (\ref{eq:mvTTmetric}) $\alpha = 1/3$.

For generic $\alpha$, the change of variables $Z= z^{1+\alpha}$ maps the complex plane into an infinitely many sheeted Riemann surface. Let us assume for simplicity (and to make contact with the vortex solutions described above) that $1+\alpha=-1/n$ with $n\in\mathbb{N}$, i.e. for the sinh-Gordon vortex $n=2$ while for the Tzitzeica vortex $n=3$. In this case the change of variables that flattens the metric is simply given by $z= Z^n$ where $Z$ is a complex coordinate on the orbifold $\mathbb{C} / \mathbb{Z}_n$, i.e. $Z\in \mathbb{C}$ and $Z\sim e^{ 2 \pi i / n} Z$ is an equivalence relation.
Note that the origin $Z=0$ has a non-trivial isotropy group under the orbifolding group $\mathbb{Z}_n$, this is precisely the reason why vortex solutions located at this special singular point have different properties from standard vortices on smooth manifolds \cite{Baptista:2012ds} and need to be treated separately.

\begin{center}
\begin{figure}[t]
\centering{
\includegraphics[scale=0.4]{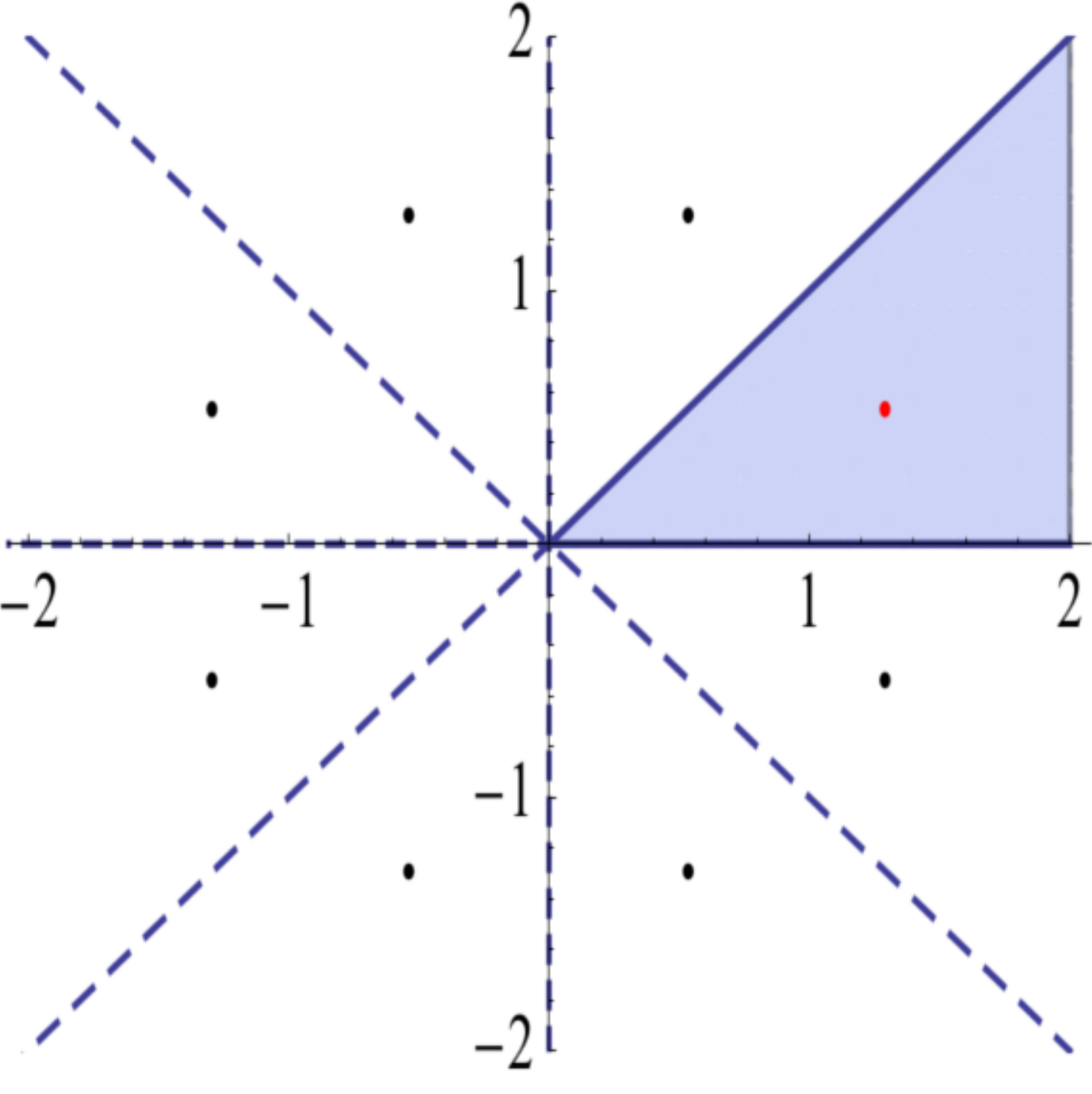}
\caption{A vortex in the cone $\mathbb{C}/\mathbb{Z}_8$, embedded in $\mathbb{C}$, and all its images under the orbifolding group $\mathbb{Z}_8$.}
\label{Figure:cone}}
\end{figure}
\end{center}

We can easily unfold the orbifold by considering $n$ copies of the original manifold $\mathbb{C} / \mathbb{Z}_n$, modulo the identification $Z\sim e^{ 2 \pi i / n} Z$; in Figure \ref{Figure:cone} we show this unfolding for the orbifold $\mathbb{C}/\mathbb{Z}_8$. For the multi sinh-Gordon and multi Tzitzeica case the change of variables that flattens the metric at the origin is given by $z= Z^{2/3}$ and $z= Z^{3/4}$ respectively. When the change of variables takes the form $z= Z^{n/m}$, with general $n,m\in\mathbb{N}^*$ the unfolding of the cone can still be performed on a multi-sheeted Riemann surface with finitely many sheets, although a pictorial description of the unfolding of the cone in this case would get rather messy. 

As we can see from Figure \ref{Figure:cone}, to find a vortex solutions when the centre is located in the interior of the cone, away from the singularity, we can simply embed the cone in flat $\mathbb{C}$ and then look for a solution with centres located at the finitely many images under the orbifolding group action $\mathbb{Z}_n$ of the original vortex location. In this way, the solution on $\mathbb{C}$ has manifest $\mathbb{Z}_n$ symmetry and yield a solution on $\mathbb{C}/\mathbb{Z}_n$.
It is clear now that, when the vortex centre coincide with the tip of the cone, the situation becomes more subtle because all the images under the orbifolding group $\mathbb{Z}_n$ degenerate to a single point with non-trivial isotropy group.

Let us write the second Bogomolny equation in $Z=z^{1/n}$ coordinates, which can be obtained from the pull-back of (\ref{vortex2}):
\begin{equation}\label{BogeqZ}
\partial_{\bar Z} \phi-i{A}_{\bar Z}\phi=0,
\end{equation}
where ${A}_{\bar Z}=n \bar z^{1-1/n}A_{\bar z}$ is the anti-holomorphic component of the connection 1-form in $Z$ coordinate. It is clear from the definitions that the gauge and Higgs fields satisfy strict periodicity
\begin{equation}\label{sym}
A_{\bar Z}(Z e^{\frac{2\pi i}{n}})=A_{\bar Z}(Z)e^{\frac{2\pi i}{n}} \;\; \text{ and } \;\;\phi(Ze^{ \frac{2\pi i}{n}})=\phi(Z).
\end{equation}

At first glance one could say that the condition (\ref{sym}) is too restrictive for a gauge field theory and that periodicity should be respected up to a gauge transformation. However, (\ref{sym}) is a consequence of the fact that we are seeking vortices on the cone and not on its $n$-covering. This is what it means to start from the equation $(\ref{vortex2})$ and not directly from (\ref{BogeqZ}). Strict periodicity is necessary to obtain integer vortex numbers, as we show below. 
On the orbifold $\mathbb{C}/\mathbb{Z}_n$ we can impose that (\ref{sym}) hold only up to a constant gauge transformation, in this way one can construct solutions with fractional vortex numbers stuck at the conical singularity \cite{Kimura:2011wh}.

While fractional vortices are necessarily fixed at the origin, integral vortices can move around and they possess a moduli space of solutions. We stress that, even if the background manifold has a conical singularity, the vortex moduli space is still a K\"ahler manifold with a well defined metric \cite{Baptista:2012ds}. The metric on the moduli space of these singular vortices is not known explicitly. The gaussian curvature has generically a delta function singularity at the tip of the cone and this prevents us from using, in a straightforward way, the expansion for the moduli space metric of vortices moving on surfaces of small curvature, obtained in \cite{Dorigoni:2013fwa}. On the other hand, as we have just shown, once we unfold the orbifold into $n$ copies living in $\mathbb C$, we obtain a smooth background manifold, even at the origin, then it is conceivable that the moduli space metric could be studied, maybe numerically, starting from $n$ vortices moving on the smooth, unfolded cone.

A simple modification of a well-known result of Jaffe and Taubes \cite{Jaffe:1980mj} (c.f. Proposition 5.1 in Chapter III) allows us to show that if $A_{\bar Z}$ and $\phi$ form a smooth solution to (\ref{BogeqZ}), then the Higgs field can be written, close to a vortex position $Z_k$,  as
\begin{equation}\label{HiggsZ}
\phi(Z)=(Z-Z_k)^{N_k} \varphi_k(Z),
\end{equation}
where $N_k\in\mathbb N^*$ and $\varphi_k$ is $C^\infty$ and non-vanishing in a neighbourhood of $Z_k$. Furthermore, when $Z_k=0$, the smooth function $\varphi_k$ is invariant under $Z\mapsto Z e^{\frac{2\pi i}{n}}$ hence, for a vortex at the origin, periodicity of $\phi$ implies that $N_k\in n\mathbb N$ and the actual vortex number (i.e. the winding of the phase $\chi$ of the Higgs field) turns out to be $\frac{N_k}{n}$ once we go back to the original coordinates $z=Z^n$. This is not surprising, since, as we can easily see from Figure \ref{Figure:cone}, every neighbourhood of the origin in  $\mathbb{C}$ is an $n$-covering of the region $Z\sim0$ in the orbifold $\mathbb{C}/\mathbb{Z}_n$, and the same vortex is counted $n$ times. When we relax (\ref{sym}) and assume periodicity only up to a gauge transformation, the above argument ceases to hold and it is possible to find fractionalized vortex solutions \cite{Kimura:2011wh}.

We can use (\ref{HiggsZ}) to expand $h = \ln \vert \phi \vert^2$ around $Z_k$, and we find that our original expansion (\ref{eq:hexp}) still holds also on conically singular spaces once we use the right set of coordinates $Z$:
\begin{equation}\label{eq:hZ}
h(Z,\bar Z)=2N_k \ln|Z-Z_k|+a(Z_k,\bar{Z}_k)+\bar b(Z_k,\bar{Z}_k) \left(Z-Z_k\right)+b(Z_k,\bar{Z}_k) \left(\bar Z-\bar{Z}_k\right)+\cdots.
\end{equation}
Of course, we have assumed in equation real analyticity of $\varphi_k(Z)$, as it should be possible to prove by following similar steps of \cite{Jaffe:1980mj}.

Since $h$ should be invariant under $Z\mapsto Ze^{i\frac{2\pi}{n}}$ and $Z_k\mapsto Z_ke^{i\frac{2\pi}{n}}$, we have $a(Z_k e^{i\frac{2\pi j}{n}})=a(Z_k)$ and $b(Z_ke^{i\frac{2\pi j}{n}})=b(Z_k)e^{i\frac{2\pi j}{n}}$ for $j=0,...,n-1$ , where we have omitted the $\bar{Z}_k$ dependence. If we consider the positions of the $n$ vortices on the $n$-covering of the cone as in figure \ref{Figure:cone}, namely, $\{Z_k e^{i\frac{2\pi j}{n}} \}_{j=0,...,n-1}$, we have
$$
\sum_{j=0}^{n-1}b(Z_k e^{i\frac{2\pi j}{n}})=\sum_{j=0}^{n-1}b(Z_k)e^{i\frac{2\pi j}{n}}=0,
$$
which characterises vortices meeting at the origin as $z_k\to 0$ \cite{MantonSut}.

From the expansion (\ref{eq:hZ}), when we set $Z_k=0$ and the vortex sits at the origin of the orbifold, we recover (\ref{hdevelop}) where the correct radial variable to use is $\rho=\vert Z\vert=r^{1/n}$ and the actual vortex number $N=N_k/n\in\mathbb N^*$, generalising the particular expansions (\ref{hsmallr}) and (\ref{hsmallrTT}) found for the sinh-Gordon and Tzitzeica vortices. Explicitly, by denoting $\rho = \vert Z\vert=r^{1/n}$, 
\begin{equation*}
h(\rho)\sim 2 n\,N \log \rho +a -\frac{\Omega(0)}{4} \rho^2 +O(\rho^4)\,,
\end{equation*} 
where the conformal factor $\Omega(0)= \frac{1}{(1+\alpha)^2}=n^2$ can be read off from (\ref{eq:metricCone}).
Translating this asymptotic form back to the original variable $r = \rho^{n}$, with $r=\vert z \vert$, we get that close to $r\sim 0 $ the vortex takes the form
\begin{equation}
h(r) \sim 2 N \log r + a - \frac{n^2}{4} r^{2/n} + O(r^{4/n})\,,\label{eq:Puiseux}
\end{equation}
matching exactly the leading asymptotic forms discussed previously for the sinh-Gordon vortex (\ref{hsmallr}), $n=2$, and the Tzitzeica vortex (\ref{hsmallrTT}), $n=3$.
Note that generically, $\vert \phi \vert^2$ is not a $C^\infty$ function of $r$, as expected from the work of Baptista on singular vortices  \cite{Baptista:2012ds}.

A similar discussion holds for the multi-vortex case. As described above, to obtain the expansion (\ref{hdevelop}), the variable to use is $\rho = r^{m/n}$, the conformal factor (\ref{eq:metricCone}) at the origin takes the form $\Omega(0)= n^2/m^2$ and the vortex profile function, written in the original coordinate $r$, can be expanded as
\begin{equation}
h(r) \sim 2 N \log r + a - \frac{n^2}{4m^2} r^{2 m /n} + O(r^{4 m/n})\,,
\label{eq:mvCone}
\end{equation}
matching precisely the leading terms in the multi sinh-Gordon case, $m/n = 3/2$ (\ref{eq:mvSG}), and the multi Tzitzeica case (\ref{eq:mvTT}), $m/n=4/3$.

It is worth mentioning that all the conformal factors used previously, i.e. in the (multi) sinh-Gordon vortex and in the (multi) Tzitzeica vortex, are not exactly of the form (\ref{eq:metricConer}) used in this Section.
The reason is simple, the conformal factor is of the form 
\begin{equation*}
\Omega(r) = e^{ \alpha h(r)} \sim r^{2\alpha} \left( a+O(r^\beta) \right)\,,
\end{equation*}
where $a$ is a non-vanishing constant and the exponent $\beta>0$ depends on the particular problem under consideration, for instance $\beta = 1$ in the (multi) sinh-Gordon case, while $\beta=2/3$ in the (multi) Tzitzeica case.
The change of variables $\rho = r^{1+\alpha}$ flattens only the leading order $r^{2\alpha}$ of the metric. For this reason the corrections to the vortex profile function (\ref{eq:mvCone}) will not be generically of order $O(r^{4 m/n})$ but rather $O(r^{\beta+2 m/n})$, i.e.
$O(r^4)$ in the multi sinh-Gordon case (\ref{eq:mvSG}) and $O(r^{10/3})$ in the multi Tzitzeica case (\ref{eq:mvTT}).

\section{Vortices on the hyperbolic cone}\label{secVorticesH}

In this section we first rederive the construction of explicit, radially symmetric vortex solutions on the entire Poincar\'e disk. Then, we take the ${\mathbb Z}_n$ orbifold of the Poincar\'e disk, so that the origin becomes a conical singularity and the base manifold becomes a ``hyperbolic cone". On this particular conically singular space we can obtain an explicit expression for the Higgs field following the discussion of Section \ref{SecCone}.

\subsection{Vortices on the Poincar\'e disk}\label{PoincareDisc}

It is convenient to define $u$ such that the conformal factor of the metric is $\Omega=e^u$.
If $u$ satisfies the Liouville's equation 
\begin{equation}\label{Liouville1}
\Delta_0 u=e^u, 
\end{equation}
then the Gauss curvature of this surface is $K=-\frac{1}{2}\Delta_0 \log \Omega = -\frac{e^{-u}}{2}\Delta_0 u=-\frac{1}{2}$. 
In this case, the Taubes equation (\ref{Taubeseq}) reduces to another Liouville's equation
\begin{equation}\label{Liouville2}
\Delta(u+h)=e^{u+h}.
\end{equation}
Therefore, vortices on surfaces of constant curvature $K=-\frac{1}{2}$ can be obtained by solving two Liouville's equations and, using the integrability properties of Liouville's equation, we can construct explicit vortex solutions on the entire surface.

We begin with the study of solutions to (\ref{Liouville1}).
\subsubsection{Liouville's equation}

The Liouville's equation $\Delta_0 \psi=e^\psi$, where $\psi$ is a real function on $\mathbb R^2$, has $3$ types of solutions \cite{Popov1993}
\begin{align}
\psi_{(1)}&=\ln\left[\frac{2(v_x^2+v_y^2)}{v^2} \right] \label{sol1} \\
\psi_{(2)}&=\ln\left[\frac{2(v_x^2+v_y^2)}{\sinh^2v} \right] \label{sol2} \\
\psi_{(3)}&=\ln\left[\frac{2(v_x^2+v_y^2)}{\sin^2v} \right], \label{sol3}
\end{align}
where the function $v$ satisfies Laplace's equation $\Delta_0 v=0$.

Let us restrict to the radially symmetric case $v=v(r)$ so that $\Delta_0 v=\frac{1}{r}(v^\prime r)^\prime=0$ implies
\begin{equation}\label{SolLaplaceR}
v(r)=C\ln(r)+D,
\end{equation}
where $C,D\in\mathbb R$ are constants. Then the three types of solutions (\ref{sol1}--\ref{sol3}) become, respectively,

\begin{align}
\psi_{(1)}&=\ln\left[\frac{2C^2}{r^2(C\ln(r)+D)^2} \right] \label{sol1bis} \\
\psi_{(2)}&=\ln\left[\frac{2C^2}{r^2\sinh^2(C\ln(r)+D)} \right] \label{sol2bis} \\
\psi_{(3)}&=\ln\left[\frac{2C^2}{r^2\sin^2(C\ln(r)+D)} \right]. \label{sol3bis}
\end{align}

The initial conditions will uniquely fix the type of solution: $\psi_{(1)},\,\psi_{(2)}$ or $\psi_{(3)}$.
In fact, the initial data set includes $3$ disjoint subsets, each one of them corresponding to one solution among (\ref{sol1bis}--\ref{sol3bis}). To see this, let us impose some general initial conditions to $\psi$. Let $r_0>0$, $v_0,v_1\in\mathbb R$, and suppose
\begin{align}\label{inconds}
\psi(r_0)&=v_0 \nonumber \\ 
\psi^\prime(r_0)&=v_1.
\end{align}
When substituted  in (\ref{sol1bis}--\ref{sol3bis}), conditions (\ref{inconds}) impose the following constraints, respectively\-:
\begin{align*}
2e^{-v_0}\left(\frac{v_1}{2}+\frac{1}{r_0} \right)^2&=1 \\
2e^{-v_0}\left(\frac{v_1}{2}+\frac{1}{r_0} \right)^2&=\cosh^2(C\ln(r_0)+D)>1 \\
2e^{-v_0}\left(\frac{v_1}{2}+\frac{1}{r_0} \right)^2&=\cos^2(C\ln(r_0)+D)<1. \\
\end{align*}

Note that the two inequalities must be strict otherwise $\psi(r_0)$ is not well-defined. This means that, given the initial data (\ref{inconds}), depending on the quantity $2e^{-v_0}\left(\frac{v_1}{2}+\frac{1}{r_0} \right)^2-1$ being zero, positive or negative, we will respectively select the solution (\ref{sol1bis}), (\ref{sol2bis}) or (\ref{sol3bis}).

\subsubsection{Vortex solutions on the conifold from the Liouville's equation}

If $u(r)$ takes one of the forms (\ref{sol1bis})--(\ref{sol3bis}), we can choose coordinates to set $C=1$ and $D=0$, and the surface $\Sigma$ with conformal factor $\Omega = e^u$ is, respectively, the once-punctured disk, the hyperbolic unit disk and a hyperbolic annulus. The only surface over which we can perform a $\mathbb{Z}_{n}$ orbifold and consider a vortex located at the origin is the hyperbolic unit disk, since it is the only surface containing this point. Vortices on hyperbolic surfaces constructed from holomorphic maps have been studied in \cite{Manton2010}\footnote{Our two Liouville's equations (\ref{Liouville1}--\ref{Liouville2}) correspond to equations (3.2) and (3.5) of this reference.}.

We then choose the solution (\ref{sol2bis}), $u(r)=\ln\left[\frac{8}{(1-r^2)^2}\right]$, yielding the metric of the hyperbolic disk $ds^2=\frac{8}{(1-r^2)^2}(dr^2+r^2d\theta^2)$. Vortices on this surface should satisfy the boundary conditions $h(r)\sim2N\ln r+\mbox{const.}+O(r^2)$ for $r\to 0$, where $N$ is the vortex number, and $h(r)\to 0$ for $r\to 1$. The only solution obtained through our method and consistent with the above boundary condition is given by $h(r)=\ln\left[\frac{2A^2}{r^2\sinh^2(A\ln r+B)} \right]-u(r)$, with $A=N+1$ and $B=0$. Hence
$$
h=2 \ln\left[\frac{(N+1) r^N(1-r^2)}{1-r^{2(N+1)}} \right],
$$   
corresponds to a multi-vortex solution located at the origin of the Poincar\'e disk, with any vortex number $N>0$.

Applying the results of Section \ref{SecCone}, we deduce that 
$$
h=2 \ln\left[\frac{(N+1) r^{N/n}(1-r^{2/n})}{1-r^{2(N+1)/n}} \right]
$$
is a solution to the Taubes equation on the Poincar\'e disk with a conical singularity at the origin and metric
$$
ds^2=\frac{8 r^{2/n-2}}{n^2(1-r^{2/n})^2}\left(dr^2+r^2d\theta^2\right)=\frac{8}{(1-\rho^2)^2}\left(d\rho^2+\frac{\rho^2}{n^2}d\theta^2\right),
$$
where $\rho=r^{1/n}$. One can check by direct calculation, that this function $h(r)$ is indeed an exact solution to Taubes equation on the hyperbolic cone with the above form for the metric.

For the Higgs field to be uniquely valued on the orbifold, we need $N\in n\mathbb N$ and the solution gives $\frac{N}{n}$ vortices at the origin. If $N$ is not an integer multiple of $n$ or if $n$ is non-integer, the solution gives a non-integer vortex number solution at the expense that the Higgs field is defined on the universal cover of the orbifold, i.e., it is periodic up to a gauge transformation.

\begin{figure}[h]
\centering{
\includegraphics[scale=0.25]{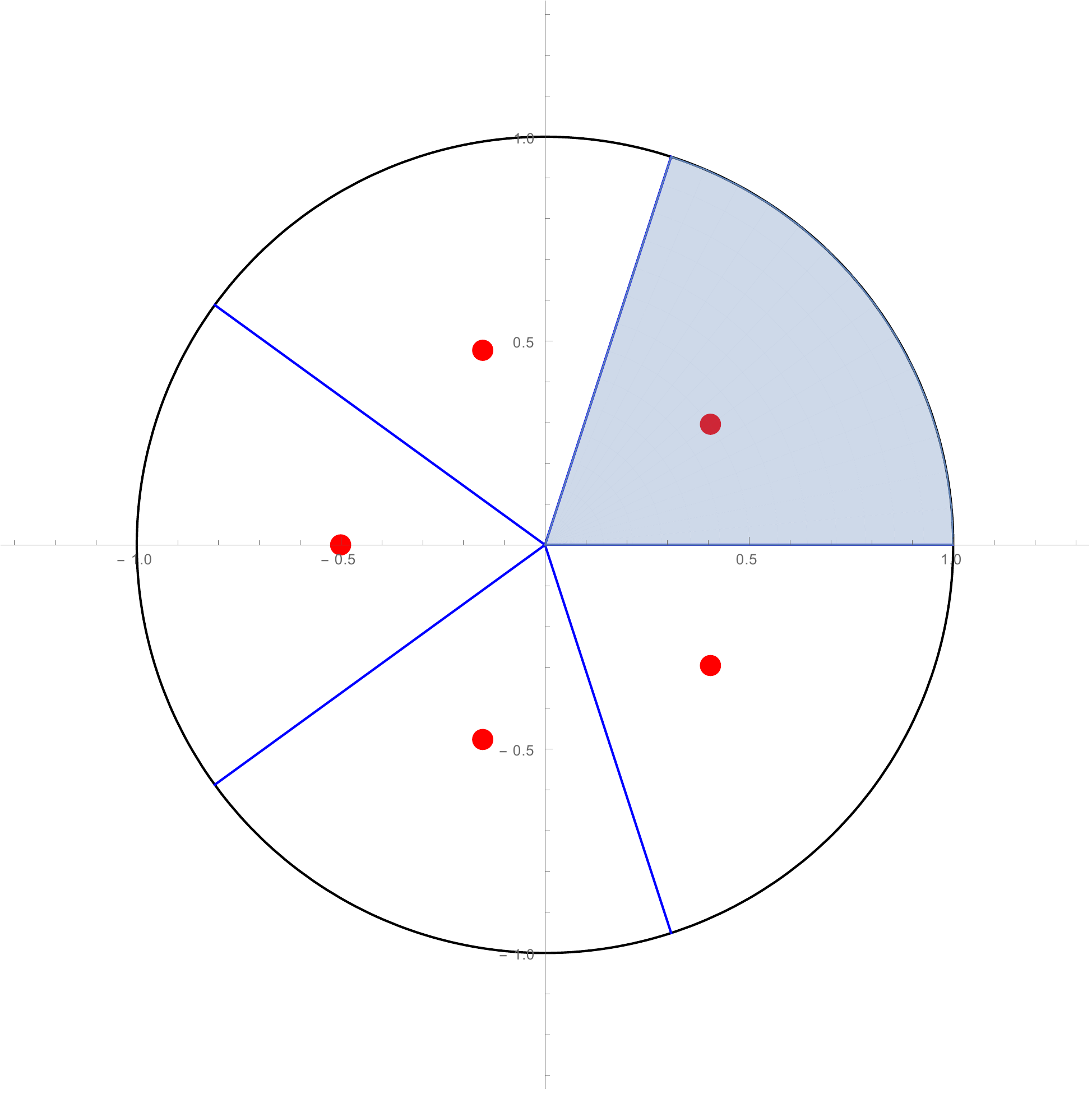}$\quad\quad$
\includegraphics[scale=0.3]{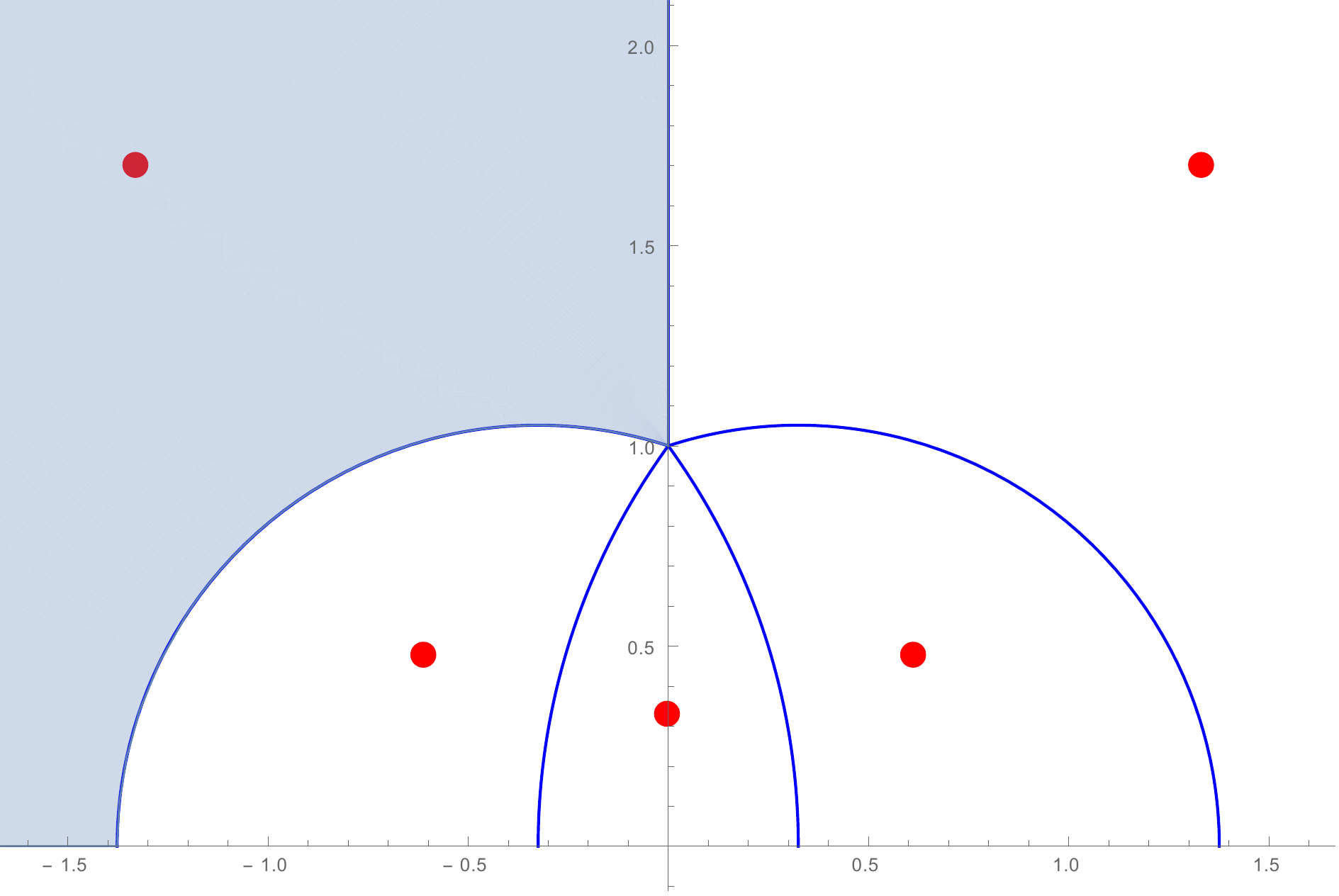}
\caption{Unfolding of the $\mathbb{Z}_5$ hyperbolic cone in the Poincar\'e disk (left) and in the upper-half plane (right). The images of the vortex centre under the $\mathbb{Z}_5$ orbifolding group are represented as red dots.}
\label{Figure:HypCone}}
\end{figure}

We can easily map this solution from the Poincar\'e disk $D=\{w\in \mathbb{C} \,\vert\, \vert w \vert\leq 1\}$ to the upper-half plane $H=\{z\in\mathbb{C}\,\vert\, \mbox{Im}\,z\geq 0\}$ by means of the M\"obius transformation
$$
w= \frac{z-i}{z+i}\,,
$$
which maps the origin of the Poincar\'e disk $w=0$ to the point $z=i$ in the upper-half plane.
The orbifold group action $\mathbb{Z}_n$ in the Poincar\'e disk geometry takes the form $w\sim e^{2 \pi i /n} w$, while in the upper-half plane it becomes $z \sim (az +b)/(cz +d)$ where $\mathbb{Z}_{n}$ is realized as a subgroup of the natural $SL(2,\mathbb{R})$ action on $H$, generated by the elements
\begin{equation*}
A_{k} = \left( \begin{matrix}
a & b\\
c& d\end{matrix}\right) =-\left( \begin{matrix}
\cos (\pi k / n) & \sin (\pi k/n)\\
-\sin (\pi k /n) & \cos (\pi k /n) \end{matrix}\right) \,.
\end{equation*}

As depicted in Figure \ref{Figure:HypCone}, we can unfold the hyperbolic cone in the Poincar\'e disk or equivalently in the upper-half plane. To find a vortex solution with centre in the interior of the cone we can simply take the explicit solution in the Poincar\'e disk or upper-half plane with centres located at the images, under the action of the orbifolding group, of the original vortex location. Since the origin $w=0$ of the Poincar\'e disk, or similarly the point $z=i$ in the upper-half plane, is a fixed point of the orbifolding group, when we insert a vortex exactly at the conical singularity all the images collapse to the vortex centre and we have to rely on the analysis carried out before.

\section{Vortices and Yang-Mills instantons}\label{SecYM}

Historically, the first multi-instanton solutions were found by Witten \cite{Witten1977}, who sought $SO(3)$-equivariant solutions to the self-dual Yang-Mills equations (SDYM) on $\mathbb R^4$. $SO(3)$-equivariance means invariance up to gauge transformations of the connection 1-form under 3-dimensional rotations acting with $2$-dimensional orbits, we will precise this notion below. As we will describe in this Section, there is a close relation between cylindrically symmetric instantons and Abelian vortices. This ``equivariance" was called ``cylindrical symmetry" by Witten and since then it has been generalized in many different ways to give rise to a large variety of Abelian and non-Abelian vortices in two and higher  dimensions \cite{Popov2009,Dolan:2010ur,Dorigoni:2014ela}.

We shall denote the 2-sphere of radius $R$ by $S^2\sim\mathbb{CP}^1$ and write its metric $g_2$ in complex and spherical coordinates
$$
g_{2}=\frac{4R^4}{(R^2+y\bar y)^2}dy d\bar y=R^2(d\theta^2+\sin^2\theta d\varphi^2),
$$
where
$$
y=R\tan\left(\frac{\theta}{2}\right)e^{-i\varphi}, \;\;\; \bar y=R\tan\left(\frac{\theta}{2}\right)e^{i\varphi},\;\;\; 0\leq\theta<\pi, \;\;\; 0\leq\varphi<2\pi.
$$

Let $\mathcal E\to M$ be a rank-2 complex vector bundle over $M=\Sigma\times S^2$, where $\Sigma$ is a Riemann surface, and $\mathcal A$ be an $SO(3)$-equivariant $\mathfrak{su}(2)$-valued connection on $\mathcal E$. It means that, under the action of $SO(3)$, $\mathcal A$ is invariant up to a gauge transformation: for any $R\in SO(3)$, there exists a matrix valued function $g_R\in SU(2)$ such that
\begin{equation}\label{SymmetryCondition}
R_{ji}{\mathcal{A}}_j(R\boldsymbol x)=g_R(\boldsymbol x){\mathcal{A}}_i(\boldsymbol x)g_R(\boldsymbol x)^{-1}-\partial_i g_R(\boldsymbol x) g_R(\boldsymbol x)^{-1}.
\end{equation}
The left hand side are the components of the pullback 1-form $R^*{\mathcal{A}}$. The group $SO(3)$ acts trivially on $\Sigma$ and through its left action on $S^2$ (for $\mathcal R\in SO(3)$, $\boldsymbol x\in S^2 \mapsto R\boldsymbol x$). 

Cylindrical symmetry imposes the following explicit form to the connection \cite{Popov2009,Manton1980,Popov2006}
\begin{equation}\label{AConnection}
\mathcal A=\left( \begin{array}{cc} \frac{1}{2}A\otimes 1+1\otimes b & \frac{1}{2}\phi\otimes\bar\beta \\
-\frac{1}{2}\bar\phi\otimes\beta &  -\frac{1}{2}A\otimes 1-1\otimes b\end{array}\right), 
\end{equation}
where $A=-i(A_z dz+A_{\bar z}d\bar z)$ is an (Abelian) $U(1)$ connection on a Hermitian complex line (rank-1) bundle $\mathcal L_1$ over $\Sigma$, $\phi$ is a section of this bundle, $b$ is the monopole connection over a complex line bundle $\mathcal L_2$ over $S^2$ given by
$$
b=\frac{1}{2(R^2+y\bar y)}(\bar y dy-yd\bar y)\,,
$$
and finally
$$
\beta=\frac{\sqrt 2 R^2}{R^2+y\bar y}dy \;\;\;\mbox{ and }\;\;\; \bar\beta=\frac{\sqrt 2 R^2}{R^2+y\bar y}d\bar y
$$
are differential forms on $S^2$.

The connection 1-form in (\ref{AConnection}) is a matrix whose entries are 1-forms. The tensor products in this expression can be regarded as the usual multiplication between scalar functions and differential forms.  


Explicitly, the components of $\mathcal A=\mathcal A_z dz+\mathcal A_{\bar z}d\bar z+\mathcal A_y dy+\mathcal A_{\bar y}d\bar y$ are

\begin{align}\label{FYM}
\mathcal A_{z}&=-i\frac{1}{2}A_{z}\sigma_3\;, \!\;\;&\!\;\; \mathcal A_{\bar z}&=-i\frac{1}{2}A_{\bar z}\sigma_3\; ,  \nonumber\\
\mathcal A_{y}&=\frac{\bar y}{2(R^2+y\bar y)}\sigma_3-\frac{R^2}{R^2+y\bar y}\frac{\bar\phi}{\sqrt 2}\sigma_-\; ,  \!\;\;&\!\;\;
\mathcal A_{\bar y}&=-\frac{y}{2(R^2+y\bar y)}\sigma_3+\frac{R^2}{R^2+y\bar y}\frac{\phi}{\sqrt 2}\sigma_+\; , \nonumber\\
\end{align}
where
$$
\sigma_3=\left(\begin{array}{cc} 1 & 0 \\ 0  & -1 \end{array} \right)\,,\;\;\;\; \sigma_+=\left(\begin{array}{cc} 0 & 1 \\ 0  & 0 \end{array} \right)\,,\;\;\;\; \sigma_-=\left(\begin{array}{cc} 0 & 0 \\ 1  & 0 \end{array} \right).
$$

The curvature 2-form $\mathcal F$ of $\mathcal A$ is 
$$
\mathcal F=d\mathcal A+\mathcal A\wedge\mathcal A=\left( \begin{array}{cc} \frac{1}{2}F-\frac{1}{2}\left(\frac{1}{R^2}-\frac{\phi\bar\phi}{2} \right)\beta\wedge\bar\beta & \frac{1}{2}(d\phi-iA\phi)\wedge\bar\beta \\
-\frac{1}{2}(d\bar\phi+iA\bar\phi)\wedge\beta & -\frac{1}{2}F+\frac{1}{2}\left(\frac{1}{R^2}-\frac{\phi\bar\phi}{2} \right)\beta\wedge\bar\beta \end{array}\right )
$$
with non-vanishing components
\begin{align}\label{FYM1}
\mathcal F_{z\bar z}&=\frac{1}{2}F_{z\bar z}\sigma_3\,, \;\;\;&\;\;\; \mathcal F_{y\bar y}&=-\frac{R^4}{(R^2+y\bar y)^2}\left(\frac{1}{R^2}-\frac{\phi\bar\phi}{2}\right)\sigma_3 \,, \nonumber\\
\mathcal F_{\bar z\bar y}&=\frac{1}{\sqrt 2}\frac{R^2}{R^2+y\bar y}(\dzbar \phi-iA_{\bar z}\phi)\sigma_+\,,  \;\;\;&\;\;\
\mathcal F_{z\bar y}&=\frac{1}{\sqrt 2}\frac{R^2}{R^2+y\bar y}(\dz \phi-iA_{z}\phi)\sigma_+\,, \nonumber\\
\mathcal F_{zy}&=-\frac{1}{\sqrt 2}\frac{R^2}{R^2+y\bar y}(\dz \bar\phi+iA_{z}\bar\phi)\sigma_- \,,\;\;\;&\;\;\ \mathcal F_{\bar z y}&=-\frac{1}{\sqrt 2}\frac{R^2}{R^2+y\bar y}(\dzbar \bar\phi+iA_{\bar z}\bar\phi)\sigma_-\,,
\end{align}
where $F=dA=F_{z\bar z}\,dz\wedge d\bar z=-i(\dz A_{\bar z}-\dzbar A_{z})dz\wedge d\bar z$.

The Hodge operator $*$ is defined by $*\mathcal F_{\mu\nu}=\frac{\sqrt{\lvert \det g\rvert}}{2}\epsilon_{\sigma\eta\mu\nu}g^{\sigma\alpha}g^{\eta\beta}\mathcal F_{\alpha\beta}$, where $g$ is the metric of the background. By applying the Hodge operator we verify that $\mathcal F$ is self-dual, i.e, $*\mathcal F=\mathcal F$, if and only if $\phi$ and $A$ satisfies the vortex equations on $\Sigma$

\begin{align*}
&2F_{z \bar z}=\frac{\Omega}{2}\left (\frac{2}{R^2}-\phi\bar\phi \right) \nonumber\\
&D_{\bar z}\phi=0,
\end{align*}
which are equivalent to (\ref{vortex1}--\ref{vortex2}) when $R=\sqrt 2$, since $B=2F_{z \bar z}$.

The conclusion is that the SDYM equations on $M=\Sigma\times S^2$ are reduced to vortex equations on $\Sigma$ once we impose an $SO(3)$ symmetry on the field. Similarly the anti-self-dual Yang-Mills equations $*\mathcal F=-\mathcal F$ can be reduced to anti-vortex equations. Conversely, given a(n) (anti-)vortex on $\Sigma$, it can be lifted to a cylindrically symmetric Yang-Mills (anti-)instanton on $\Sigma\times S^2$.

\subsection{Instantons from the sinh-Gordon and Tzitzeica vortices}

In this Section we derive a solution of the SDYM equations from the vortices described in Sections \ref{SecSGVortex}, \ref{SecTTVortex} and \ref{sec:mvSGTT}. 
Not to overcrowd this Section with too many equations, we will give explicit formulas only for the sinh-Gordon vortex, similar results can be derived in a straightforward manner for the Tzitzeica case as well as for the multi-vortex case.

From our sinh-Gordon vortex solution $h_{sG}$ we can reconstruct the Higgs field $\phi=e^{h_{sG}/2+i\chi}$, where $\chi$ is a real function defined on each open patch depending on the gauge choice, while using the Bogomolny equation (\ref{vortex2}) we can obtain the gauge field from the Higgs field: $A_{\bar{z}} = - i\partial_{\bar{z}} \log \phi=-i \partial_{\bar{z}} h_{sG}/2 +\partial_{\bar{z}} \chi$.
Close to the origin we can simply use the asymptotic expansion (\ref{hsmallr}) and obtain
\begin{align}
\phi(z,\bar{z}) &\sim \beta_{sG}^2 \vert z \vert \,e^{i\chi}( 1 + O (\vert z \vert^2))\,,\\
A_{\bar{z}}(z,\bar{z}) &\sim -\frac{i}{2}\frac{z}{z\bar{z}} \left(1+\frac{\vert z\vert }{2\beta_{sG}^2} +O(\vert z \vert^2)\right)+\partial_{\bar{z}}\chi.
\end{align}

We can use (\ref{FYM}) to uplift this vortex solution to an instanton solution on $\Sigma\times S^2$, provided that $R=\sqrt{2}$, and the components of the connection for the instanton solution close to $z\sim 0$ are
\begin{align*}
\mathcal A_{z}&=\left(\frac{1}{4}\frac{\bar z}{z\bar z}-\frac{1}{8\beta_{sG}^2}\frac{\bar z}{\lvert z \rvert}-\frac{i}{2}\dz\chi \right)\sigma_3 \,,\!\;\;&\!\;\; \mathcal A_{\bar z}&=\left(-\frac{1}{4}\frac{\bar z}{z\bar z}+\frac{1}{8\beta_{sG}^2}\frac{\bar z}{\lvert z \rvert}-\frac{i}{2}\dzbar\chi \right)\sigma_3\,,  \nonumber\\
\mathcal A_{y}&=\frac{\bar y}{2(2+y\bar y)}\sigma_3-\frac{\sqrt{2}}{2+y\bar y}\lvert z \rvert \beta_{sG}^2\, e^{-i\chi}\,\sigma_-\,,  \!\;\;&\!\;\;
\mathcal A_{\bar y}&=-\frac{y}{2(2+y\bar y)}\sigma_3+\frac{\sqrt{2}}{2+y\bar y}\lvert z \rvert \beta_{sG}^2\, e^{i\chi}\,\sigma_+ \,,\nonumber
\end{align*}
where we omitted higher terms in $\vert z\vert$.

The components of the curvature two-form can be easily calculated and the only non-vanishing components are
\begin{align}
\mathcal F_{z\bar z}&=\frac{1}{8\beta_{sG}^2\lvert z \rvert}\sigma_3, \;\;\;&\;\;\; \mathcal F_{y\bar y}&=-\frac{2}{(2+y\bar y)^2}\sigma_3, \nonumber \\
\mathcal F_{\bar z y}&=-\frac{\sqrt 2}{2+y\bar y}\left( \beta_{sG}^2\frac{\lvert z \rvert}{\bar z}-\frac{z}{2} \right)e^{-i\chi}\sigma_-,  \;\;\;&\;\;\
\mathcal F_{z\bar y}&=\frac{\sqrt 2}{2+y\bar y}\left( \beta_{sG}^2\frac{\lvert z \rvert}{z}-\frac{\bar z}{2} \right)e^{i\chi}\sigma_+, \nonumber ,
\end{align}
where we neglected, once again, higher terms in $\lvert z \rvert$. 

The corresponding expansion for the instanton connection $\mathcal{A}$ and field strength $\mathcal F$ as $z\to\infty$ can be calculated in the same way from the expansion as $r\to\infty$ for $\phi$ and $A$ given by (\ref{hlarger}).

We want to check now that the uplifting of our sinh-Gordon vortex does indeed correspond to a $1-$instanton solution on $\Sigma\times S^2$.
The instanton number $N$ is defined as the integral
$$
N=-\int_{\Sigma\times S^2}C_2,
$$
where
$$
C_2=\frac{1}{8\pi^2}\left(\tr(\mathcal F\wedge\mathcal F)-\tr\mathcal F\wedge\tr\mathcal F\right)=d\left(\frac{1}{8\pi^2}\tr(d\mathcal A\wedge\mathcal A+\frac{2}{3}\mathcal A\wedge\mathcal A\wedge\mathcal A) \right)
$$
is the second Chern form. The wedge product $\wedge$ between matrices indicates the usual multiplication of matrices but applying the wedge product between the entries.

It is easy to check that the Chern forms splits in the product of the first Chern class for the vortex field strength and the first Chern class for the monopole connection, so that the instanton number can be rewritten as
\begin{equation}
N= -\int_{\Sigma\times S^2} C_2 = \frac{i}{2\pi} \int_\Sigma F\cdot \frac{i}{2\pi} \int_{S^2} db\,.
\end{equation}
Since the sinh-Gordon vortex has vortex number one and the monopole has magnetic charge one, it follows that the instanton number is also $N=1$. This means that the SDYM solution on $\Sigma\times S^2$ obtained from the uplifting on the sinh-Gordon vortex on $\Sigma$ corresponds precisely to a $1-$instanton located at the origin of $\Sigma$ and spread along the $S^2$.

Note that even if $\mathcal{F}$ is singular close to $z\sim 0$, the instanton number is still finite and integer. This follows from the Bogomolny equation (\ref{vortex1}): the sinh-Gordon (and Tzitzeica) vortex solution has a diverging magnetic field close to the origin of $\Sigma$ because of the diverging conformal factor $\Omega$, nonetheless this singularity is integrable and the vortex has a finite and quantised magnetic flux. 

Similar results can be derived from the Tzitzeica vortex, furthermore, higher instanton number solutions can be obtained by uplifting in a similar fashion our multi-vortex solutions of Section \ref{sec:mvSGTT}. 

\section{Conclusion}

In this paper we first reviewed the construction of Abelian vortex solutions from the sinh-Gordon equation and the elliptic Tzitzeica equation.
These solutions are not known in explicit forms over the entire background but only in the asymptotic regimes $r\to 0$ and $r\to \infty$.

Using the non-linear superposition rule described by Baptista, we constructed multi-vortex solutions on top of the sinh-Gordon and the Tzitzeica vortex and analysed their properties with various numerical simulations.
The vortices constructed with this procedure are all defined on surfaces with conical singularities, and, for this reason, the usual expansion for the vortex profile function (\ref{eq:hexp}) ceases to hold. 

For this reason we analysed the problem of finding vortex solutions on conically singular spaces and we showed that with a careful change of coordinates (from $z$ to $Z$), for which the metric becomes flat and the cone can be unfolded in the complex plane, the problem reduces simply to the study of vortex solutions invariant under the action of an orbifold symmetry. 
In particular we see from equation (\ref{eq:hZ}) that, close to the vortex location, the Higgs field $\vert \phi \vert^2$ is real analytic in the new coordinates $Z$. 

When the vortex is located away from the conical singularity, the solution close to the vortex centre is smooth in the original coordinates $z$ as well. On the contrary, when the vortex is located precisely at the conical singularity, the asymptotic form of the profile function (\ref{eq:Puiseux}), when expressed in the original coordinates, takes the form of a Puiseux series in $|z|^{2/n}$, which means that generically the Higgs field $\vert\phi\vert^2$ is only $C^0$ as a function of $r=\vert z\vert$, as already expected from Theorem 2.1 of \cite{Baptista:2012ds}. However, we note from equation (\ref{hsmallr}) that, for the sinh-Gordon vortex, $\vert\phi\vert^2$ is $C^\infty$ as a function of the original coordinate $r$. 

As an additional example of our analysis, we discuss the Taubes equation on the Poincar\'e disk. On this surface, explicit vortex solutions can be obtained from the Liouville equation and from them we can analytically construct vortex solutions on the hyperbolic cone. Once we map the Poincar\'e disk to the upper half plane $H$, the orbifold group can be realized as a discrete subgroup of the natural $SL(2,\mathbb{R})$ acting on $H$. When uplifted to $4$-dimensions, vortex solutions on the upper-half plane $H$ become instantons with cylindrical symmetry on $\mathbb{R}^4 \sim H\times S^2$, it would be interesting to understand what kind of instantons can be obtained by uplifting these vortex solutions on $H/ \mathbb{Z}_n$ to $4$-dimensions.

It would also be interesting to apply our approach to vortices on conifolds to the case of compact surfaces, for example, a sphere with one or more conical singularities \cite{EtoFuji2009}, to analyse the effect of the orbifold action and the compact nature of the background on the global properties of the vortex.

In the final part of our work, we described how to uplift our multi-vortex solutions, defined on the conically singular surface $\Sigma$, to instanton solutions, with cylindrical symmetry, on the background $M=\Sigma\times S^2$ with a product metric. The four dimensional manifold $M$ is K\"ahler with K\"ahler form 
\begin{equation}
\omega=i\Omega\, dz\wedge d\bar z+i\frac{16}{(2+y\bar y)^2}dy\wedge d\bar y.
\end{equation}
Moreover, this metric has non-vanishing scalar curvature and therefore has no self-dual or anti-self-dual Weyl tensor (c.f. Proposition 10.2.2 of \cite{MDbook}). In this case, the 6-dimensional twistor space of $M$ has a non-integrable almost complex structure \cite{Boyer86,LeBrun86} and the Penrose-Ward transform does not apply. These solutions go beyond the analysis of integrable (anti-)self-dual backgrounds, in which case the vortex equations would arise as compatibility conditions of linear differential equations on the twistor space.

\section*{Acknowledgements}
We are grateful to Maciej Dunajski, Nick Manton, Norman Rink and Alexander Cockburn for useful discussions. F.C. is grateful for the support of Cambridge Commonwealth, European $\&$ International Trust and CAPES Foundation Grant Proc. BEX 13656/13-9. D.D. is grateful for the support of European Research Council Advanced Grant No. 247252, Properties and Applications of the Gauge/Gravity Correspondence.

\section*{Appendix. Numerical Analysis }
\setcounter{equation}{0}
\appendix
\def\theequation{\thesection{A}\arabic{equation}}

\label{sec:app}

The asymptotic forms for the solution of the Taubes equation (\ref{Taubeseqdelta}) can be fixed analytically only in radial reductions of the sinh-Gordon, $\Omega= e^{-h_{sG}/2}$, or the Tzitzeica case, $\Omega = e^{-2h_{TT}/3}$, by exploiting the Painlev\'e property of the two ODEs.
Unfortunately for a generic metric no such methods exist and if we want to compute multi-vortex solutions to (\ref{system2}) or (\ref{system2TT}), we have to rely on a numerical calculation. 

Furthermore, when we apply our superposition rule for vortices, we need to use the modified conformal factors $\tilde \Omega = e^{h_{sG}/2}$ or $\tilde \Omega = e^{h_{TT}/3}$, but we do not have the explicit solutions to the sinh-Gordon and Tzitzeica vortices for all values of $r$, so we will have to obtain numerically the sinh-Gordon and Tzitzeica vortex solutions interpolating between the two known asymptotic forms.
This problem and all the subsequent studies for multi-vortex solutions have been solved numerically in the following way, first instead of working for $r\in \mathbb{R}$ we cut away the $r\rightarrow \infty$ and the singular point $r\sim 0$ by working with $r \in [\epsilon,R]$ and checking that the solution does not change as we send $\epsilon \rightarrow 0$ and $R\rightarrow \infty$.

Secondly instead of working with the profile function $h$ it is better to strip away the $\log-$like singularity by working with:
\begin{equation*}
h(r)= u(r)+2 N \log (r/R)\,,
\end{equation*}
in this way the $\delta$ function on the right-hand side of Taubes equation disappears and $u$ satisfies:
\begin{equation*}
\nabla^2 u +\Omega\left( 1- \frac{r^{2N}}{R^{2N}} e^u\right)=0\,.
\end{equation*}
From the asymptotics of $h$ we can read those of $u$: $u(\epsilon) \sim a +2 \log R + O(\epsilon^\alpha)$, where $\alpha>0$ depends on the particular metric at hand, while for $r\sim R$ the $\log$ term that we added vanishes (but not its derivative) so $h(R) = u(R) \sim \Lambda e^{-R}$ (remember that all our metrics are asymptotically flat, i.e. $\Omega\to 1$ as $r\to\infty$).
To obtain a solution for $u$ we implemented both a shooting and a cooling method and the two solutions coincide within numerical errors.

Let us first construct the sinh-Gordon and Tzitzeica vortices. We know from Sections \ref{SecSGVortex}-\ref{SecTTVortex} that the vortex number can only be $N=1$ and we need to solve for
\begin{align}
\frac{d^2 u_{sG}}{d r^2}+\frac{1}{r}\frac{d u_{sG}}{dr} +\frac{e^{-u_{sG}/2}}{r} \left( 1- \frac{r^{2}}{R^{2}} e^{u_{sG}}\right)&=0\,,\\
\frac{d^2 u_{TT}}{d r^2}+\frac{1}{r}\frac{d u_{TT}}{dr} +\frac{e^{-2u_{TT}/3}}{r^{4/3}} \left( 1- \frac{r^{2}}{R^{2}} e^{u_{TT}}\right)&=0\,,
\end{align}
with boundary conditions
\begin{align}
u_{sG}(\epsilon) &= 4 \log \beta_{sG}+2 \log R - \frac{\epsilon}{\beta_{sG}^2} + O(\epsilon^2)\,,\\
u_{sG} (R) & = \Lambda_{sG} K_0(R)+O(e^{-2 R})\,,
\end{align}
and similarly 
\begin{align}
u_{TT}(\epsilon) &=  \beta_{TT}+2 \log R - \frac{9 e^{-2 \beta_{TT}/3 }\epsilon^{2/3}}{4} + O(\epsilon^{4/3})\,,\\
u_{TT} (R) & = \Lambda_{TT} K_0(R)+O(e^{-2 R})\,.
\end{align}
We fixed $\epsilon = 10^{-4}$ and $R=30$, so that the higher terms in the boundary conditions are numerically negligible.
\begin{center}
\begin{figure}[t]
\centering{
\includegraphics[scale=0.2]{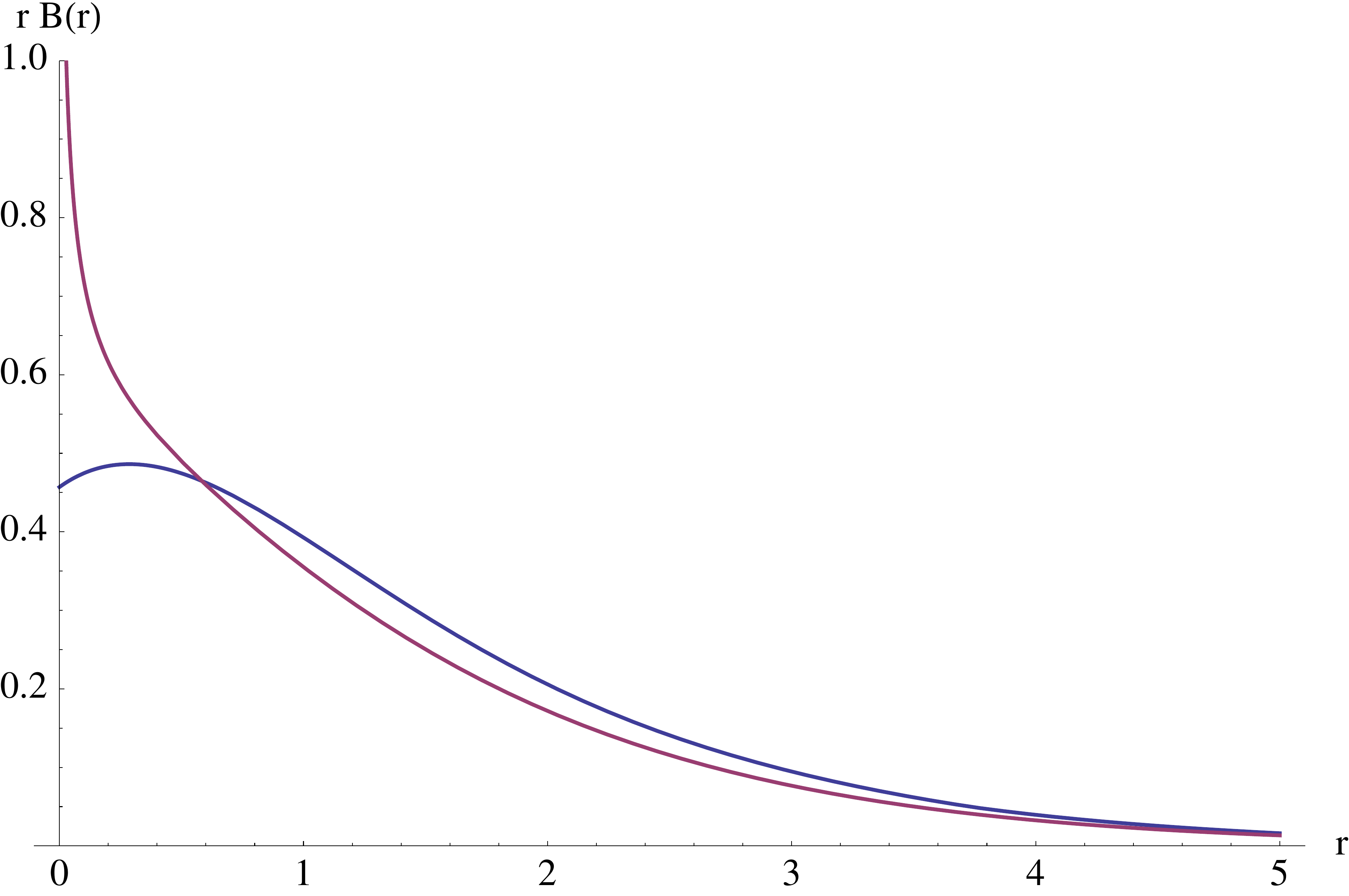}
\caption{Plot of the magnet field $B(r)$ times the radial coordinate $r$ for the sinh-Gordon vortex, finite at $r=0$, and the Tzitzeica vortex, diverging at $r=0$.}
\label{fig:rB}}
\end{figure}
\end{center}
We can see in Figure \ref{fig:rB} that the magnetic field $B(r)$, for the two numerical solutions, is obviously localized in a region close to the origin and decays exponentially to $0$ for large $r$. We note that in the Tzitzeica case, due to the diverging conformal factor present in the Bogomolny equation (\ref{vortex1}), we have $r B(r) \sim r^{-1/3}$ close to $r\sim0$, however this singularity at the origin is integrable and the magnetic flux is actually finite and quantized.

With the sinh-Gordon and Tzitzeica vortex solutions in our hands, we are now in the position to study the multi-vortex problem.
Let us focus for simplicity on the multi sinh-Gordon problem (\ref{system2}), which translated to the $\tilde u(r) = \tilde h(r) -2 N \log (r/R)$ variable takes the form:
\begin{align}
&\label{eq:umvSG}\frac{d^2 \tilde u}{d r^2}+\frac{1}{r}\frac{d \tilde u}{dr} +e^{h_{sG}/2}  \left( 1- \frac{r^{2 N }}{R^{2N }} e^{\tilde u}\right)=0\,,\\
&\notag \tilde u(\epsilon) = \tilde a+ O(\epsilon^3)\,,\\
&\notag \tilde u'(\epsilon) = O(\epsilon^2)\,.
\end{align}
To obtain the behaviour (\ref{eq:mvSG}) of $u$ close to $r\sim0$ we require that $\tilde u(r)=\sum_{n\geq 0 } a_n r^n$, and imposing that (\ref{eq:umvSG}) is satisfied order by order, fixes all the coefficients $a_n$ in terms of $\tilde a = a_0$.
By choosing $\epsilon = 10^{-4}$ we can set $u'(\epsilon)\sim0$, so that we can perform a shooting method where the shooting parameter $\tilde a$ is chosen in such a way that $u(R) = h(R) \sim \tilde \Lambda e^{-R} \to 0$, where we set $ R= 30$.
We checked the precision of our numerical simulations by evaluating the magnetic flux as a function of the vortex number $N$ and the shooting parameter $\tilde a$.
\begin{center}
  \begin{figure}[t]
  \begin{tabular}{ l | r }
\includegraphics[scale=0.17]{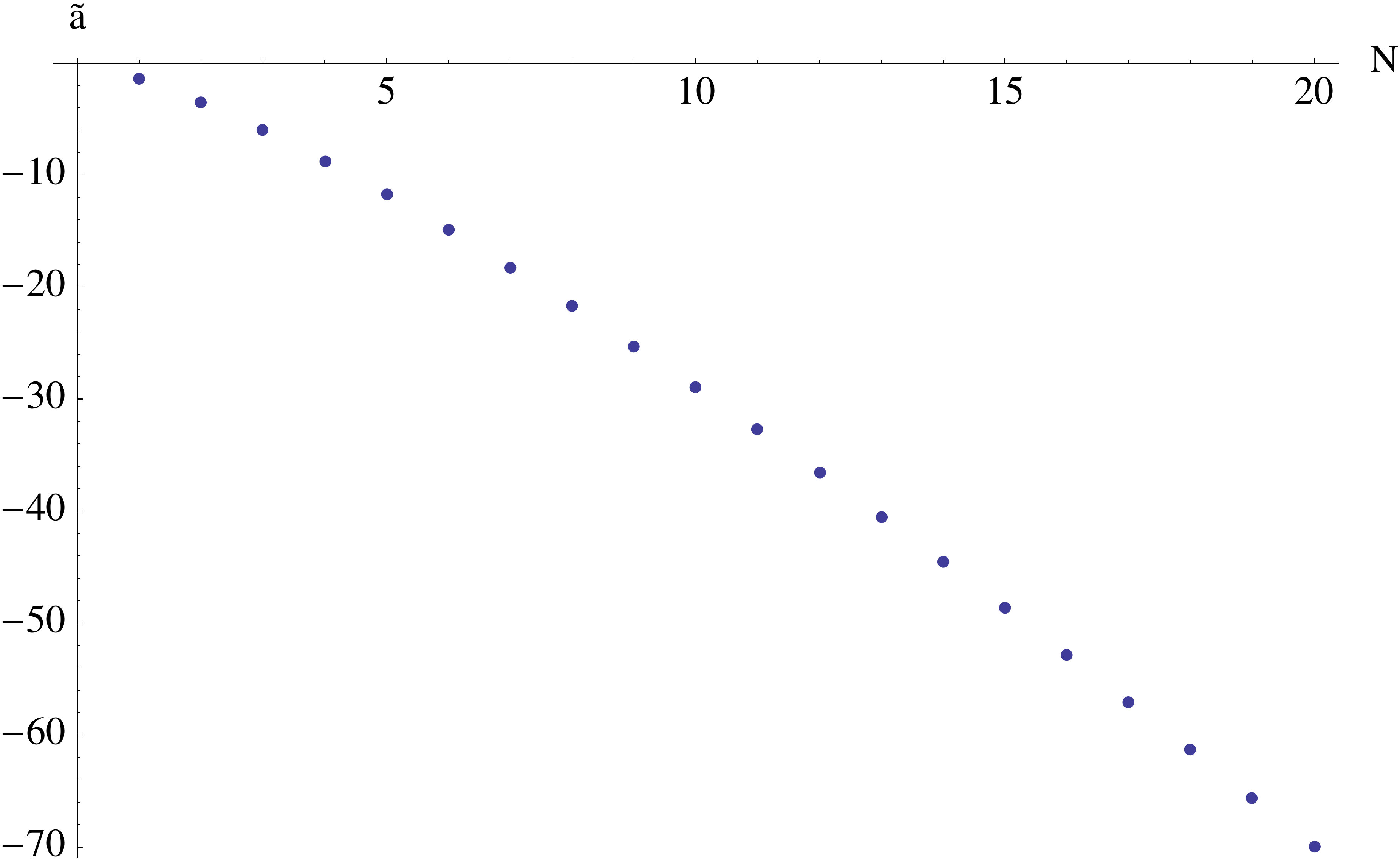} & \includegraphics[scale=0.174]{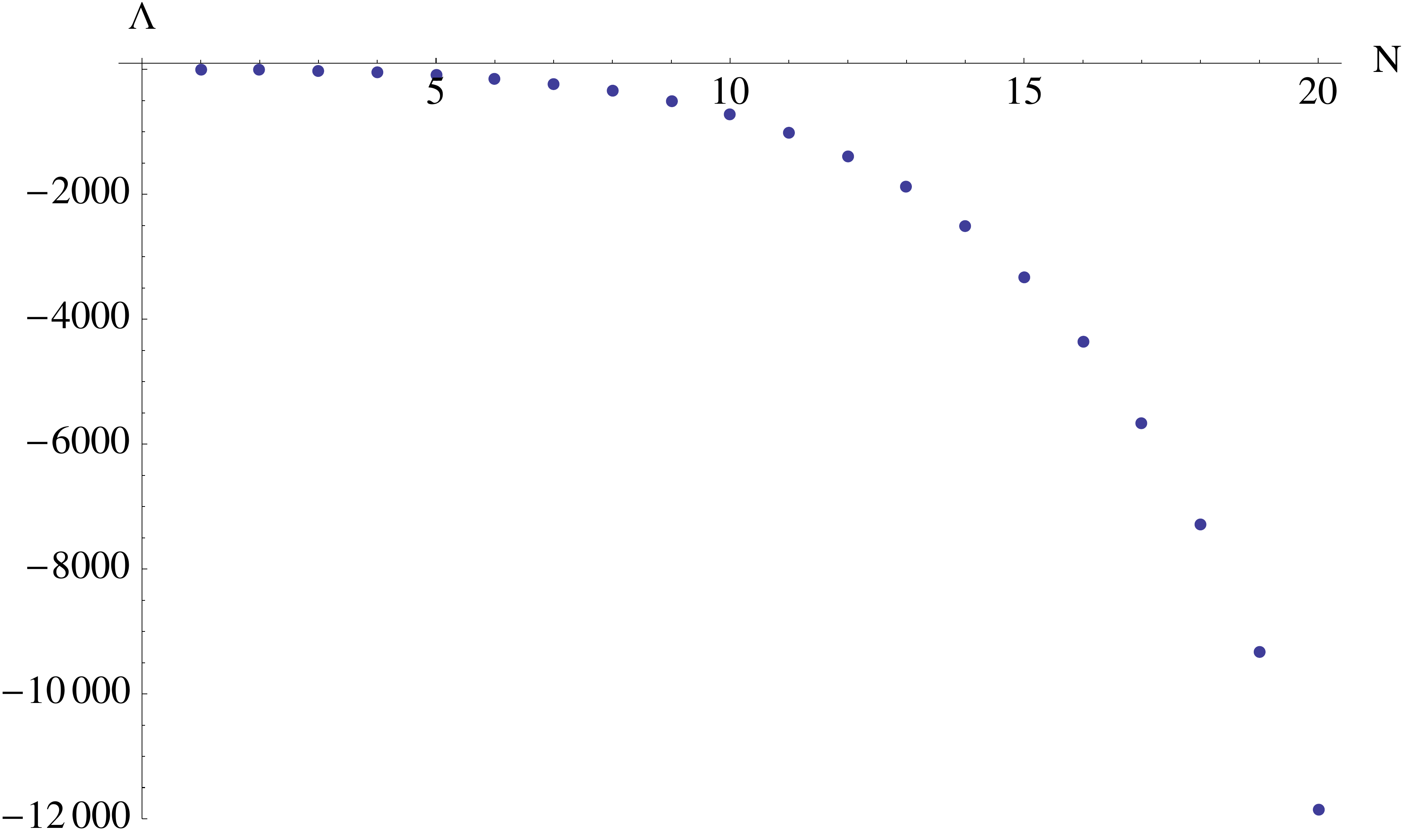}
  \end{tabular}
  \caption{Values for $\tilde{a}$, left plot, and the vortex strength $\tilde{\Lambda}$, right plot, as a function of the vortex number $N$ for the multi sinh-Gordon vortex.}
  \label{fig:SG}
  \end{figure}
\end{center}
In Figure (\ref{fig:SG}) we summarize the results of our multi sinh-Gordon vortex numerical analysis by plotting the values of $\tilde a$ and $\tilde \Lambda$ as a function of the vortex number $N\in \{1,...,\,20\}$: once the vortex number is fixed the solution is uniquely determined by $\tilde a$ or equivalently by the vortex strength $\tilde \Lambda$.

We repeated an identical numerical analysis for the multi Tzitzeica vortex and obtained similar plots for $\tilde a ,\,\tilde \Lambda$ as functions of the vortex number $N$.

\bibliography{BiblioArticle} 
\bibliographystyle{unsrt}
\end{document}